\documentclass{emulateapj}

\newcommand{\ie}{{i.e., }}

\newcommand{\simgt}{{\raise-.5ex\hbox{$\buildrel>\over\sim$}}}
\newcommand{\simlt}{{\raise-.5ex\hbox{$\buildrel<\over\sim$}}}

\newenvironment{packed_enum}{
\begin{enumerate}
  \setlength{\itemsep}{1pt}
  \setlength{\parskip}{0pt}
  \setlength{\parsep}{0pt}
}{\end{enumerate}}

\usepackage{multirow}

\slugcomment{}

\shorttitle{Surface Brightness Profiles of Dwarf Galaxies: I. Profiles and Statistics}
\shortauthors{Herrmann, Hunter, \& Elmegreen}

\begin{document}

\title{Surface Brightness Profiles of Dwarf Galaxies: I. Profiles and Statistics}

\author{Kimberly A.\,Herrmann$^{1,2}$, Deidre A.\,Hunter$^1$, and Bruce G.\,Elmegreen$^3$}
\affil{$^1$Lowell Observatory, 1400 West Mars Hill Road, Flagstaff, AZ, 86001, USA; kah259@psu.edu, dah@lowell.edu}
\affil{$^2$Current address: Penn State Mont Alto, 1 Campus Drive, Mont Alto, PA 17237, USA}
\affil{$^3$IBM T.\,J.\,Watson Research Center, 1101 Kitchawan Road, Yorktown Hts., NY, 10598, USA; bge@us.ibm.com}

\begin{abstract}
Radial surface brightness profiles of spiral galaxies are classified into three types: (I) single exponential, or the light falls off with one exponential to a break before falling off (II) more steeply, or (III) less steeply.  Profile breaks are also found in dwarf disks, but some dwarf Type~IIs are flat or increasing out to a break before falling off.  Here we re-examine the stellar disk profiles of 141 dwarfs: 96 dwarf irregulars (dIms), 26 Blue Compact Dwarfs (BCDs), and 19 Magellanic-type spirals (Sms).  We fit single, double, or even triple exponential profiles in up to 11 passbands: $GALEX$ FUV and NUV, ground-based $UBVJHK$ and H$\alpha$, and $Spitzer$ 3.6 and 4.5~$\mu$m.  We find that more luminous galaxies have brighter centers, larger inner and outer scale lengths, and break at larger radii; dwarf trends with $M_B$ extend to spirals.  However, the $V$-band break surface brightness is independent of break type, $M_B$, and Hubble type.  Dwarf Type~II and III profiles fall off similarly beyond the breaks but have different interiors and IIs break $\sim$twice as far as IIIs.  Outer Type~II and III scale lengths may have weak trends with wavelength, but pure Type~II inner scale lengths clearly $decrease$ from the FUV to visible bands whereas Type~III inner scale lengths $increase$ with redder bands.  This suggests the influence of different star formation histories on profile type, but nonetheless the break location is approximately the same in all passbands.  Dwarfs continue trends between profile and Hubble types such that later-type galaxies have more Type~II but fewer Type~I and III profiles than early-type spirals.  BCDs and Sms are over-represented as Types~III and II, respectively, compared to dIms.
\end{abstract}


\section{INTRODUCTION}
\begin{center}
Look at a galaxy!  Its disk light \\
Falls exponentially- is that right? \\
If you look deeply, often you'll see \\
Signs of us- in both Types II and III! \\
Why do we exist?  Explore the gas, \\
Motions near and far.  Profile the mass. \\
Search with care; do whatever it takes. \\
We are \underline{Surface Brightness Profile Breaks}!
\end{center}

As the riddle indicates, to first order galactic disk light falls off exponentially with radius.  This trend was observed early on \citep{p1940}.  Within a few decades, it was discovered that not all spirals are a simple exponential plus a central bulge \citep{f1970}; some edge-on spirals appeared to be sharply truncated at large radius (van der Kruit 1979; van der Kruit \& Shostak 1982; Shostak \& van der Kruit 1984; van der Kruit 1987; de Grijs et al.\,2001; Kregel et al.\,2002; Pohlen et al.\,2002; Kregel \& van der Kruit 2004).  More recently, large deep studies (Erwin et al.\,2005; Pohlen \& Trujillo 2006 (PT06); Erwin et al.\,2008 (EPB08); Guti\'{e}rrez et al.\,2011 (GEAB11)) have shown breaks are common in the surface brightness exponential fall-off in spirals.  Type~I profiles follow a single exponential to the full probed depth; these are surprisingly rare.  The majority of late-type spirals now appear to be ``truncated'' (Type~II) where the disk light continues to fall off, but more steeply beyond the break (PT06).   Lastly, many spirals have ``anti-truncated'' (Type~III) profiles where the fall off is {\it shallower} beyond the break (Erwin et~al.\,2005).   Figure~5 of PT06 shows examples of Type~I, II, and III spirals.

\begin{figure*}
\plotone{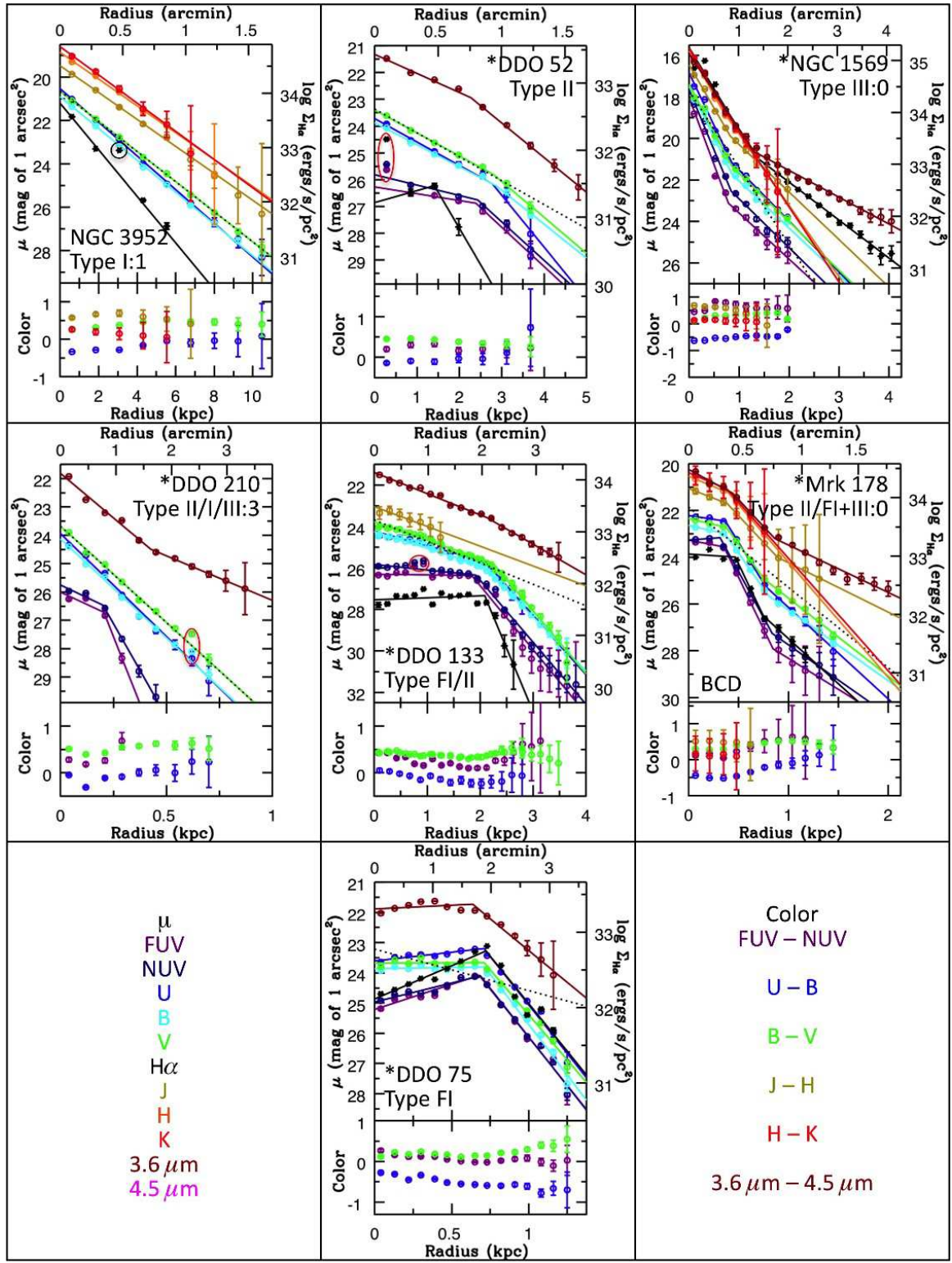}
\caption{Surface brightness profiles and radial color distributions for seven representative profile types, defined by the ensemble of passbands for each galaxy.  {\it Upper left:} Type I, single exponential. {\it Central column:} Type II, truncated, {\it (top)} pure Type II, {\it (center)} FI/II combination where the profile in at least one band is FI (flat or increasing out to a break) whereas at least one other band has a II profile, and {\it (bottom)} pure FI.  {\it Upper right:} Type III, anti-truncated.  {\it Center left:} an example where all the major profile types are represented (such that types are listed generally blue to red for multiple types).  {\it Center right:} an example galaxy best fit by two breaks.  Note that:  (1) the dotted black line is the best Type I fit to the $V$-band data, (2) LITTLE THINGS dwarfs are indicated with an asterisk preceding the name, (3) eliminated data points are circled such that red indicates all contained points are eliminated whereas black or blue indicate only H$\alpha$ or $U$ data are eliminated, respectively, and (4) numbers after a colon indicate the following: 0 = all forced to be broken, 1 = all forced to be single, 3 = at least one forced to be single whereas others forced to be broken.  See the online AJ version for similar plots for the full sample of 141 galaxies.  The $UV$ data are in AB magnitudes whereas the other broadband data are on the Vega system.  \label{Rep7} }
\end{figure*}

Why there are three different radial profile types (I, II, and III) is still a mystery, including why light falls off as an exponential at all.  Ideas proposed to explain why there are three different profile types include angular momentum limits in a collapsing protogalactic cloud \citep{vdk87} or star formation (SF) thresholds caused by radial variations in gas density or phase \citep{k89,s04,eh06,b+10}.  One recent development deals with radial color trends.  Bakos et~al.\,(2008) found that Type~II spirals have redder colors beyond the break; their interpretation is that interactions with spiral arms cause older, redder stars to migrate outwards.  Additionally, the color changes appear to be due to differing stellar populations such that the mass-to-light ratio varies, so Type~II breaks exist in the light distribution but are much less conspicuous in the mass profile.  Some simulations further support this interpretation \citep{r+08a,r+08b,ms+09}, although simulations by \citet{sb+09} indicate that a disk-metallicity gradient and an inside-out formation scenario may also be important.  Additionally, Younger et~al.\,(2007) have suggested that Type~III profiles may generally be caused by minor mergers.

Spirals are not the only galaxies with different stellar profile types: disk-like dwarf galaxies also have these properties.  \Citet{he06} presented broadband ($UBV$ and some $JHK$) disk surface photometry for 136 dwarfs and found that $\sim$30\% have a surface brightness break.  Additionally, dwarfs have a fourth classification: flat or even increasing intensity with increasing radius in the central regions (FI) before falling off beyond some break.  Figure~\ref{Rep7} shows example dwarf galaxy profiles of the different types.

Zhang et~al.\,(2012) performed a spectral energy distribution analysis of a subset of 34 of the above 136 dwarfs.  They found that 27 dwarfs ($\sim$80\%) have Type~II profiles corresponding to a steeper radial decline of the SF rate (SFR) in the outer areas with respect to the inner disk.  They also found that outer disk scale lengths are shorter in bluer bands (probing young stellar populations) than in redder bands (probing older stellar populations), at least for Type~II dwarfs.  They interpreted these results as an outside-in shrinking of the star-forming disks.  This is in contrast to spirals where evidence suggests an inside-out disk evolution (e.g.\,Larson 1976; Chiappini et al.\,1997; Mo et al.\,1998; Naab \& Ostriker 2006).

Breaks are not only seen in the local universe, but also at intermediate redshifts ($0.6 < z < 1.0$; P\'{e}rez 2004; Trujillo et~al.\,2009).  \Citet{atb08} studied 238 Type~II galaxies separated into three redshift bins: $0.1 < z < 0.5$, $0.5 < z < 0.8$, and $0.8 < z < 1.1$ and also compared their results to the $z = 0$ sample of PT06.  They found trends between the break radius, $R_{br}$, the surface brightness at the break, $\mu_{br}$, the inner scale length, $h_1$, and the ratio of $R_{br}/h_1$ with $M_B$ as well as trends with redshift such that Type~II galaxies at higher redshift have shorter $R_{br}$, brighter $\mu_{br}$, roughly unchanging $h_1$, and smaller $R_{br}/h_1$ than local analogs.

Dwarfs are the most common type of galaxy in the universe and the closest analogs to the building blocks of larger systems, so they are important to understand for piecing together the process of galaxy formation.  Disk-like dwarfs do not have spiral arms but still have Type~II profiles, so a radial color trend analysis similar to that of Bakos et~al.\,(2008) is needed for dwarfs.  The first step in the statistical analysis is to fit Type~I, II, and III profiles as reliably as possible to a large sample of dwarf surface brightness profiles.  Upon further examination of the original 136 profiles from \citet{he06} supplemented with five additional dwarfs as well as ancillary data at additional passbands, we noticed that more profiles than identified by Hunter \& Elmegreen may be fit as well (if not better) by broken exponentials instead of a single exponential.  Often breaks became more evident when additional passbands were examined.  Furthermore, different bands highlight different stellar populations (such as UV and H$\alpha$ for recent SF), so any changes with wavelength could be evidence of galaxy evolution.  Therefore, we undertook a re-examination of the profile types of the entire 141 dwarf sample in up to 11 passbands.

Here we present the statistical results of a human-assisted computerized fitting of the multi-wavelength data set of surface brightness profiles.  In Section 2 we review the sample and the data sources.  We explain the fitting process, including many complications, in Section 3.  Section 4 contains copious results before we end with a summary and brief discussion.  Future papers in this series will focus on topics such as radial color trends and mass profiles, break relationships to \ion{H}{1} gas surface density and kinematics (in the smaller LITTLE THINGS\footnotemark[4] 41 galaxy subsample), and an exploration of the two-dimensional (2D) images for each profile type.

\footnotetext[4]{The LITTLE (Local Irregulars That Trace Luminosity Extremes) THINGS (The HI Nearby Galaxy Survey) team is an international collaboration studying a multi-wavelength sample of 37 dIm and 4 BCD galaxies to understand SF in these small systems \citep{LTdata}.}

\section{THE DATA}
Our galaxy sample (see Table~\ref{Sample}) is from the survey of Hunter \& Elmegreen (2004), which includes 94 dwarf Irregulars (dIms), 26 Blue Compact Dwarfs (BCDs), and 20 Magellanic-type spirals (Sms)\footnotemark[5].  (The Sm F567-2 is missing from this study and two dIms, F473-V1 and F620-V3, have been added.)  BCDs are similar to dIms, but with central concentrations of gas, stars, and SF.  They are generally believed to be undergoing a burst of SF.  Sm galaxies mark the transition between late-type spirals and the irregular galaxy class.  Most of the galaxies were chosen to be relatively nearby ($<$30 Mpc), not obviously interacting, containing \ion{H}{1} gas so they could be forming stars, and with a wide range in $M_B$ and SF rate surface density.  Hunter \& Elmegreen (2004) presented narrow-band H$\alpha$ images of these galaxies obtained in 22 observing runs primarily at Lowell Observatory.  (Note that sometimes no H$\alpha$ emission was detected.)

\footnotetext[5]{Note that the term ``dwarf'' is not well defined in the literature. Hubble (1936) may have first used the term and Reaves (1952) performed the first survey for dwarf galaxies, but neither clearly defined the term. Over the years people have chosen individual definitions based most often on absolute magnitude (Tolstoy 2001). However, ``dwarf'' is now routinely used to refer to all Magellanic type irregular galaxies.  Our sample covers galaxies with absolute magnitudes comparable to the LMC and fainter. We use the term ``dwarfs'' here to refer to our whole sample of dIm, BCD, and Sm galaxies.}

Optical broad-band images of galaxies are dominated by the past 1~Gyr of SF whereas near infrared light integrates the SF over a galaxy's lifetime.  $UBV$ images from 27 observing runs are given by Hunter \& Elmegreen (2006) for 136 dwarfs and images from at least one band of $JHK$ are presented for 41 (26 dIm, 12 BCD, and 3 Sm) from an additional 9 observing runs.  (Note that four dwarfs were missing $UB$ data and three more were missing $U$ data.)  The remaining 5/141 galaxies (DDO~125, Mrk~67, NGC~1705, NGC~2101, and NGC~3109) were observed earlier with a different detector and so were not included in Hunter \& Elmegreen (2006).  Most of the $UBVJHK$ images were obtained with the Lowell Observatory 1.1~m or 1.8~m telescopes.  

\begin{deluxetable*}{ccccccccc}
\tabletypesize{\scriptsize}
\tablecaption{Galaxy Sample\label{Sample}}
\tablewidth{0pt}
\tablehead{\colhead{Galaxy} & \colhead{Other Names\tablenotemark{a}} & \colhead{Group\tablenotemark{b}} & \colhead{$D$(Mpc)} & \colhead{Ref\tablenotemark{c}} & \colhead{$E(B-V)$\tablenotemark{d}} & \colhead{$M_B$\tablenotemark{e}} & \colhead{Type\tablenotemark{f}} & \colhead{Notes\tablenotemark{g}}}
\startdata
0467-074 & 2dFGRS N293Z129, ISI 96 1032-0121                & Im &	107.9 & \dots & 0.03 & -18.31 & II & \dots	\\
1397-049 & ISI 96 1051+0227                                                      & Im &	118.7 & \dots & 0.02 & -17.23 & II & \dots	\\
A1004+10 & PGC 29428, UGC 5456, IRAS F10046+1036    & Im &	6.5   & \dots   & 0.01 & -15.60 &	III & \dots	\\
A2228+33 & PGC 69019, UGC 12060, IRAS F22282+3334 & Im &	16.9  & \dots  & 0.01 & -17.24 &	III+II & B \\
CVnIdwA	 & UGCA 292                                                                 & Im &	3.6    & 1         &	0.01 & -12.16 &	FI &	LT
\enddata
\tablenotetext{}{Note: Table~1 is available in its entirety in the electronic edition of the {\it Astronomical Journal}.  A portion is shown here for guidance regarding its form and content.}
\tablenotetext{a}{Selected alternate identifications obtained from NED.}
\tablenotetext{b}{Morphological Hubble types are from de Vaucouleurs et al.\,(1991; RC3).  If no type is given there, we have used types given by NED.}
\tablenotetext{c}{Reference for the distance to the galaxy. If no reference is given, the distance was determined from $V_{\rm{GSR}}$ given by RC3 and a Hubble constant of 65~km$^{-1}$~Mpc$^{-1}$, as in \citet{he06}.  An * indicates the distance was determined from the galaxy's radial velocity, given by RC3, corrected for infall to the Virgo Cluster (Mould et al.\,2000) and a Hubble constant of 73~km$^{-1}$~Mpc$^{-1}$.  Only six galaxies have distances determined with the latter method and they are all part of the LITTLE THINGS sample (Hunter et~al.\,2012).  Note that surface photometry is independent of distance.}
\tablenotetext{d}{Foreground reddening from Burstein \& Heiles (1984).}
\tablenotetext{e}{Absolute magnitude in $B$-band, calculated from Hunter \& Elmegreen (2006).  Galaxies without $B$-band data are indicated by ``NB''.}
\tablenotetext{f}{Final profile type per galaxy determined here.  Galaxies with two breaks are listed as ``inner type + outer type'' and ones with different types depending on wavelength are generally listed blue to red.}
\tablenotetext{g}{Notes column: B = barred, PM = peculiar morphology, LT = LITTLE THINGS (Hunter et~al.\,2012) member, T = THINGS \citep{THINGS} member, VO = $V$-band only}
\tablenotetext{}{References: (1) Dalcanton et al.\,2009; (2) Karachentsev et al.\,2004; (3) Karachentsev et al.\,2003a; (4) Karachentsev et al.\,2006; (5) Dolphin et al.\,2002; (6) Sakai et al.\,2004; (7) Dolphin et al.\,2003; (8) Karachentsev et al.\,2003b; (9) Tolstoy et al.\,1995a; (10) Karachentsev et al.\,2002; (11) Makarova et al.\,1998; (12) Meschin et al.\,2009; (13) Schulte-Ladbeck et al.\,2001; (14) Sakai et al.\,1999; (15) Pietrzynski et al.\,2006; (16) Miller et al.\,2001; (17) Freedman et al.\,2001; (18) Karachentsev et al.\,1996; (19) Grocholski et al.\,2008; (20) Tolstoy et al.\,1995b; (21) Gieren et al.\,2006; (22) Momany et al.\,2002; (23) Lynds et al.\,1998; (24) M\'{e}ndez et al.\,2002; (25) Gieren et al.\,2008.}
\end{deluxetable*}

Satellite UV images were also obtained with the {\it Galaxy Evolution Explorer} ({\it GALEX}, Martin et al.\,2005) because the UV light traces SF from the past $\sim$200~Myr. {\it GALEX} produced images in two passbands: FUV, with a bandpass of $1350 - 1750$\AA, an effective wavelength of 1516\AA, and a resolution of 4\arcsec\,and NUV with a bandpass of $1750 - 2800$\AA, an effective wavelength of 2267\AA, and a resolution of 5.6\arcsec.  The FUV and NUV data are on the AB magnitude system whereas the data in all other broad bands are on the Johnson/Cousin system.  Hunter et al.\,(2010) analyzed surface photometry and derived SF rates from archival {\it GALEX} images of 44 galaxies.  (Note that NUV data were available for DDO~88, DDO~165, and DDO~180 but FUV images were not.)  To probe the extreme outer stellar disks of dIms, Hunter et al.\,(2011) obtained deep {\it GALEX} UV images of an additional four galaxies and a deeper image of DDO~53.  The LITTLE THINGS team obtained data on another 12 galaxies \citep{z+12}.

We obtained mid-infrared (3.6 and 4.5 $\mu$m) images with the Infrared Array Camera (IRAC, Fazio et al.\,2004) from the {\it Spitzer} archives to probe old stars, dust, and embedded SF. The IRAC PSF at 3.6~$\mu$m has a FWHM of 1.7\arcsec\,(Fazio et al.\, 2004). Hunter et al.\,(2006) analyzed data on 21 galaxies in our sample. Other data were taken by several {\it Spitzer} Legacy projects: The Local Volume Legacy and the Spitzer Infrared Nearby Galaxy Survey.  The surface photometry from the Legacy Projects included 3.6~$\mu$m data but not 4.5~$\mu$m, resulting in significantly more 3.6 than 4.5 $\mu$m profiles.

We were careful to remove or mask out foreground stars and background objects, and in most cases we did a 2D fit to the sky around the galaxy before subtraction.  Note that significant care was taken to prevent over- or under-subtraction of the sky to ensure that breaks are not artifacts of data handling (see Section 3.3 of Hunter \& Elmegreen 2006 as well as Hunter et al.\,2011 for a discussion of sky subtraction and comparisons between images obtained with different telescopes).  The surface photometry was azimuthally averaged by integrating in ellipses that increase in semi-major axis length in $\sim$10\arcsec\ steps.  Though these galaxies are irregular in shape, this analysis appears to work well.  (See Section 3.4 about asymmetries.)  For each galaxy, the geometric parameters (morphological center, position angle, and minor-to-major axis ratio $b/a$) were determined from the $V$-band outer isophotes and applied to all passbands the same way.  See Section 3 and Table 2 of Hunter \& Elmegreen (2004, 2006) for more details on the surface photometry including the geometric parameters and ellipse widths.  The surface brightness uncertainties were determined from Poisson statistics using galaxy and sky counts and range from $\sim$0.001 mag in the (relatively) bright, central regions to $\sim$1 mag in the fainter outskirts.  (See Equation~1 and the surrounding text in Hunter \& Elmegreen 2006 for more details.)

\section{HUMAN-ASSISTED COMPUTER BREAK FITTING}
While it is possible to fit profile breaks by eye, when there are 141 galaxies and data in up to 11 passbands, it is preferable to devise an automatic method.  More importantly, in these small and faint dwarfs with as few as four points in their radial profiles, sometimes it is difficult to decide qualitatively (1) if there should be a break and (2) where the break should occur.  Consequently, a program was written to determine the ``best'' fit, single or double exponentials, for each of the total 776 profiles.  Because exponentials in surface brightness appear as lines when the data is treated logarithmically (i.e., in magnitudes), ``broken'' or ``double'' exponentials are synonymous with two-line fits.

Single and double exponentials are not the only functions that could be used to fit surface brightness profiles: Sersic functions \citep{Sersic1,Sersic2} could also be used.  However, for a direct comparison with spirals (such as the study of Bakos et~al.\,2008), the single and double exponential fits are required.  Also, Sersic fits could naturally fit Type~I and II (down-bending) profiles, but some additional changes would need to be applied to the equations to fit Type~III (up-bending) profiles.  Lastly, cases where the inner and outer sections are precisely two exponentials, without any curvature, will not be well fit by Sersic functions.  A re-fitting of the complete sample with adjusted Sersic equations may be a project for a future paper in this series.

\subsection{Basics of the Fitting Program}
For all the linear fitting, we started with a minimizing $\chi^2$ function from \citet{NumRec}, which returns the slope, $y$-intercept, and corresponding probable uncertainties.  If $y$ standard deviations are used to weight the data, the function also returns $\chi^2$ and the goodness-of-fit probability $q$ (that the fit would have $\chi^2$ this large or larger) calculated from the incomplete gamma function.  If no weighting is applied, $q$ is returned as 1.0 and the returned value of $\chi^2$ is normalized to unit standard deviation on all points.  For each of the 141 dwarf galaxies, each passband profile is examined by the program separately.  Consequently, the program determines the best fit for each passband independently.

The first step is to determine the best Type~I linear fit of all the radial data for the galaxy and passband under consideration.  Weighting by 1/$\sigma$ is used because the original 1/$\sigma^2$ weighting gave visibly poor fits to the data because the uncertainties are not distributed via a Gaussian function, instead ranging from 0.001 to $\sim$1 mag.  Consequently, 1/$\sigma^2$ weighting would weight some data as much as $\sim10^6$ times more than others, yielding a fit very biased by the (inner) data with very low uncertainty.  To keep the different radii independent yet still account for higher uncertainties of some data, the 1/$\sigma$ weighting was used as a compromise between complete uncertainty weighting and no weighting at all.

\begin{figure}
\epsscale{1.03}
\plotone{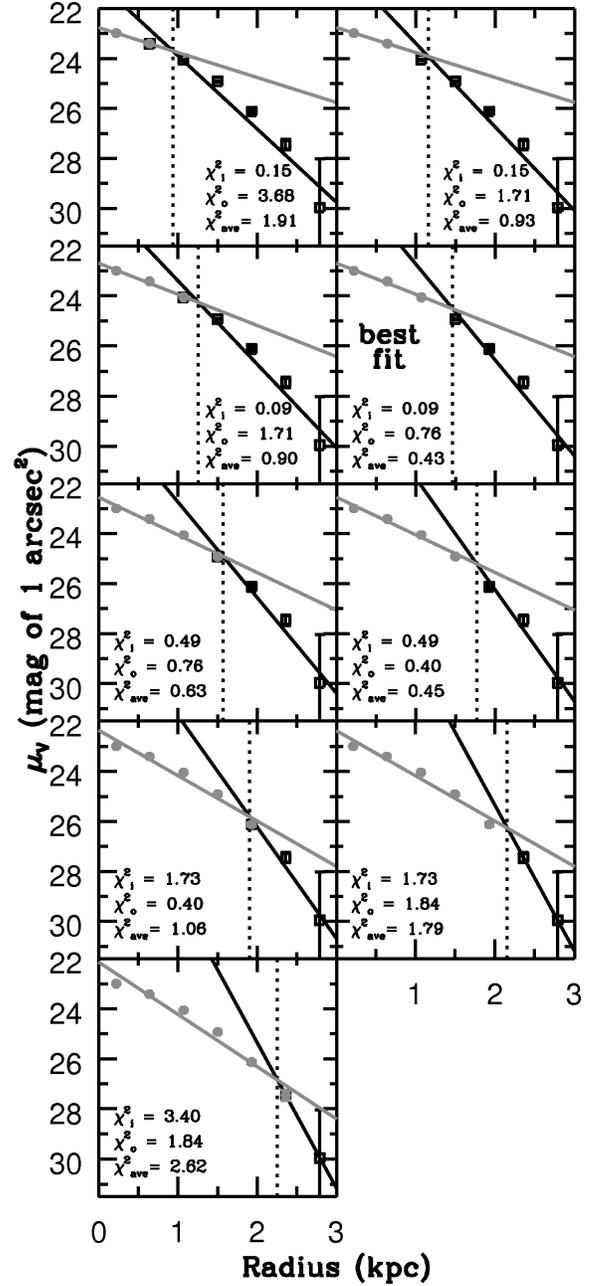}
\caption{Example of iterative fitting method for the $V$ data of DDO~43.  The data are separated into every possible pair of two subsets containing at least two points.  The inner and outer subsets are shown by gray circles and black squares, respectively.  The 1/$\sigma$ weighted best linear fits are shown by solid gray and black lines, respectively.  The vertical dotted lines show the resulting break locations.  The {\it left} panels show examples of ``knee'' cases where the last point in the inner subset is also included as the first point in the outer subset whereas the subsets of the {\it right} panels do not overlap to test the case where the break occurs firmly between two data points.  The inner, outer, and average $\chi^2$ values are given in each panel and the best fit is labeled accordingly. \label{method} }
\end{figure}

Next the program iteratively breaks the radial profile into every possible pair of subsets with at least two data points (see Section 3.2.4 about fitting only two data points) and determines the best linear fit of each subset, again using 1/$\sigma$ weighting.  Figure~\ref{method} illustrates this procedure for the $V$ data of DDO~43.  When isolating every possible pair of subsets, the program tries (1) including each inner point in both subsets (\ie  sometimes the break occurs essentially at a ``knee'' point, where that point fits well on both sides, shown in the left side of Figure~\ref{method}) as well as (2) testing each inner point on only one side (\ie  alternately the break occurs firmly in between two points, shown in the right side of Figure~\ref{method}).  The ``best'' broken fit has (1) the minimum sum of the $\chi^2$ values from the inner and outer fits such that (2) a ``break'' (\ie  intersection of the inner and outer linear fits) occurs between the correct data points.  This second test was implemented to prevent fits with very low summed $\chi^2$ values where the ``break'' occurred at unphysically negative or very large radii.

In this manner, 1/$\sigma$ weighting is used for the Type~I fitting and determining the data set separations for the break locations but no weighting is used for the final broken fits.  The uncertainties on values such as the scale length ($h_R = 2.5\log(e)/$slope), the break location ($R_{br}$= the radius where the two fits intersect), the surface brightness at the break ($\mu_{br}$), and ratios such as $R_{br}/h_{R,o}$ and $h_{R,i}/h_{R,o}$ where the $i$ and $o$ subscripts denote inner and outer sections, respectively, are determined from propagating the uncertainties on the slope and $y$-intercept values.  Uncertainty propagation is always calculated under the assumption that the variables are not correlated (\ie  uncertainties in $R_{br}$ and $h_{R,o}$ are both calculated from the slopes, but that is ignored when determining the uncertainty in $R_{br}/h_{R,o}$).  When 1/$\sigma$ weighting was used for each segment of a broken profile, the propagated uncertainties on calculated values were typically larger than the values themselves.  In the no weighting case, the uncertainties on the slope and intercept values are multiplied by a factor of the reduced $\chi^2$ values to normalize $\chi^2$ to unit standard deviation on all points in an attempt to estimate the unknown standard deviations of the data points.  Since the reduced $\chi^2$ values are typically very small, corresponding to excellent fits, this process significantly decreases the final uncertainties.  Consequently, for the final fitting of the broken segments, no weighting is used, resulting in uncertainties smaller than their calculated values.  Note, however, that these uncertainties are only lower limits since systematic uncertainties (such as zero point and distance uncertainties) have not been included.

\subsection{Complications}
Ideally, the above fitting should determine the best fit of each profile.  However, some intervention was necessary to deal with complications.

\subsubsection{Single versus Broken?}
First of all, how should the program choose whether a profile is better fit by a single exponential (Type~I) or a broken profile (Type~II, III, or FI)?  Some parameters which logically should contain useful information are the individual values of $\chi^2$, goodness-of-fit probability $q$, as well as slope similarity on both sides of the break.  We checked the values of these parameters in cases that were either clearly Type~I or had strong breaks by eye to gauge how well they would do in distinguishing cases.  Unfortunately, there was no one obvious method to discriminate between the two cases.  For the comparison between dwarfs and spirals to be meaningful, the dwarf profiles needed to be classified as single or broken as reliably as possible, so some care was taken to determine how to distinguish between these two classifications.

Many of the profiles that appeared to be well fit as Type~I by eye had $q_I > 0.95$, indicating an excellent fit, where the $I$ subscript indicates the single fit to the complete profile.  Others had $\chi^2_I < 0.8 \times 0.5(\chi^2_i + \chi^2_o)$ where the $i$ and $o$ subscripts represent the inner and outer fits of the best broken fit.  The factor of 0.8 was determined empirically; the visual inspection of the best-fit profiles preferred this value over other values.  However, it is understandable that a multiplicative factor less than 1 may be necessary.  If the data truly define a single line, the fit should get better with more data points to average out noise.  Consequently, the factor of 0.8 serves to decrease the average $\chi^2$ value from the two broken fits since each will have of order half the size of the whole sample and thus have fewer data points to constrain the fit.  Alternately, in general if the data truly define two lines, then the $\chi^2_I$ value will be large anyway.  Thus the program defines a profile to be broken unless $q_I > 0.95$ or $\chi^2_I < 0.8 \times 0.5(\chi^2_i + \chi^2_o)$.

\begin{figure}
\plotone{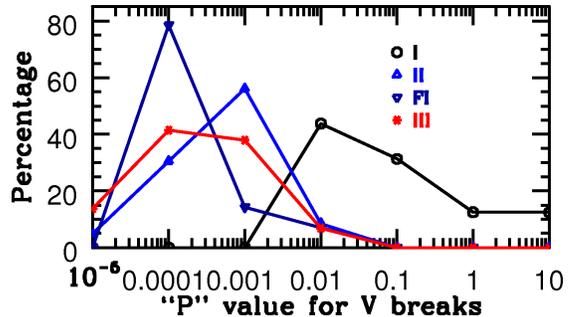} 
\caption{Distribution of $P$ values by percentage of profile types for broken and single fits for the $V$ data set where $P$ is a measure of how well a single line fits the data; large and small $P$ values correspond to single and broken profiles, respectively.  The Type~I profiles are relatively separate from the broken profiles, so the classification between broken and single fits appears to be robust. \label{Pfig}}
\end{figure}

The parameter $P = 0.8 \times 0.5(\chi^2_i + \chi^2_o)/\chi^2_I$ is a measure of the break strength such that very broken profiles should have tiny $P$ values.  Figure~\ref{Pfig} shows the distribution of $P$ by percentage of profile types for the $V$ band data set.  Since the distribution of Type~I profiles is relatively separate from those of the broken (II, III, and FI) profiles, the classification between broken and single fits appears to be robust.  A one-dimensional (1D) two sample Kolmogorov-Smirnov (K-S) test (implemented from \citet{NumRec}) yields a probability $< 10^{-5}$ that the distribution of $P$ for the Type~Is is the same as for those of the broken profiles but 29\%, 10\%, and 5.3\% chances that the III and FI, III and II, and II and FI $P$ values are drawn from the same distributions, respectively.

In retrospect, we realized that the slope difference between the two sides and the average slope uncertainty could be used to determine if a profile should be Type~I or broken.  A broken profile should have a slope difference larger than the average uncertainties whereas the reverse should be true for Type~I profiles.  Of the 16 instances of Type~I profiles in $V$, only one (Haro~8) would be broken by this test and all 125 profiles that were originally classified as being broken are still broken by this test.

\subsubsection{Forcing Some Fits to be Single or Broken}
Unfortunately, the above prescription for identifying single from broken profiles is not perfect, so human intervention was needed even after implementing the ``automatic'' way to determine between single and broken profiles described above.  The profiles of some galaxies were {\it forced} to be single, broken, or even single in at least one passband but broken in others because the program was not sophisticated enough to make the best determination in all cases.  Of the 141 galaxies, the program was allowed to choose between single or broken for 99 (70\%) of the sample.  Of the remaining 42 galaxies, 23 (16\% of the total) were forced to be broken, 4 (3\% of the total) were forced to be single, and 15 (11\% of the total) were forced to be broken in some bands but single in others.  While those percentages at first appear disturbing because they could have an effect on the percentage types discussed in Section 4.3, in every case that a profile was forced to be single or broken, information from multiple bands was used to make the decision.

\begin{figure*}
\epsscale{0.97}
\plotone{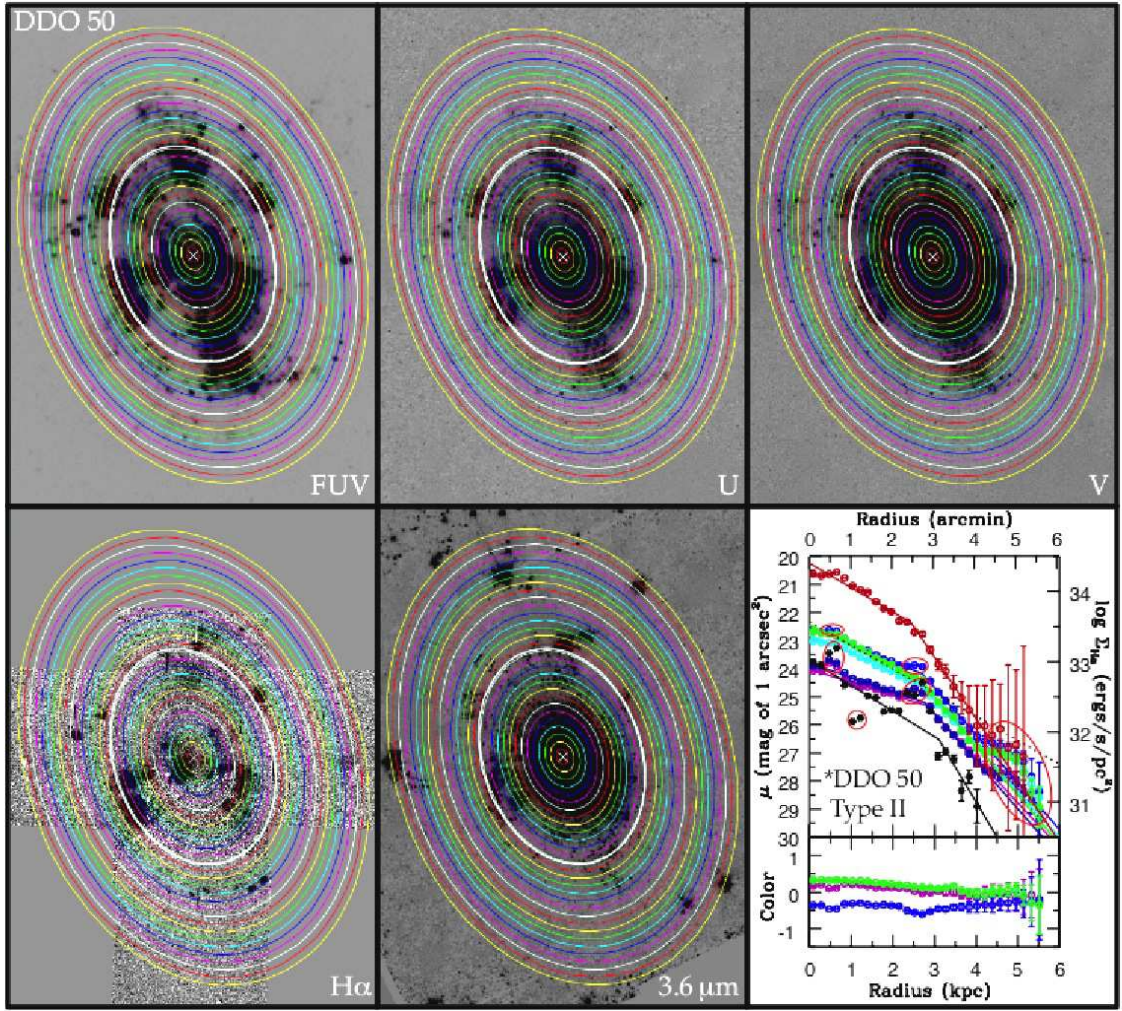}
\caption{Images of DDO~50 in most of the observed bands with the surface photometry rings superimposed to highlight some excesses from SF clumps.  The NUV and $B$ images were omitted from the figure because they are very similar to the FUV, $U$, and $V$ images.  In each image, the white X indicates the galactic center and the thick white ellipse indicates the location of the break in the $V$ band.  The lower right panel shows the azimuthally averaged surface brightness profiles with the eliminated data points enclosed by red ellipses.  See the color legends in Figure~1. Note that the continuum subtracted H$\alpha$ image is completely dominated by discrete clumps of emission and these clumps are more pronounced in the bluer bands, such as FUV, than redder bands, such as $V$.  Note also that the outer excess just North of East appears to be primarily responsible for the outer excess in the azimuthally averaged profiles. \label{DDO50} }
\end{figure*}

Out of the 23 forced breaks, 18 had clear breaks chosen by the program in most bands, but in at least one band, the break was hidden by large uncertainties, but still apparent to the eye.  The remaining five did not have particularly large uncertainties but did have program-assigned breaks in most bands and nearby visible breaks in the remaining bands, so all were just forced to be broken, though the program still chose where the break should occur in each passband separately.  These profiles were forced to be broken essentially to increase the samples of breaks in poorly represented passbands.  The four that were forced to be unbroken appeared single in all four to seven bands, but the program was fooled by a little noise in a few bands.  Of the remaining 15 that were forced to be broken in some bands but single in others, six had at least one band that should be single (by the program's standards, either because it should be Type~I or because the data in that band did not go deeply enough to probe the break area of other bands) but profiles in at least three passbands that should be broken (by the program's standards) and at least one additional band where the break was again hidden by large uncertainties, yet still visible.  In those six cases the breaks that were hidden from the program by large uncertainties were forced to be broken.  Two more had clear breaks in all the passbands except one band that was not deep enough and thus was just forced to be single.  The last seven had some clear breaks but at least some data, mostly in H$\alpha$, that really should have been Type~I.

\subsubsection{Eliminating Outlying Points from the Fitting} 
Some hand-trimming of outlying points was needed in 53 galaxies (38\%) primarily due to deviations from SF clumps, especially in H$\alpha$ and the bluer bands.  A large subsample of the images with eliminated data was inspected visually to confirm that SF clumps are generally responsible for the eliminated outlying data.  (Figure~\ref{DDO50} shows an example galaxy with many eliminated points.)

In 20/53 (16\% of the 129 with detectable H$\alpha$), only H$\alpha$ outliers were eliminated because of at least one of the following: (1) the same trend was visible in the H$\alpha$ data as in additional bands on both sides of the questionable point, (2) there was no clear large scale trend in the H$\alpha$ data, but some small scale trends were consistent with those in other passbands whereas eliminated H$\alpha$ points were not, (3) eliminated H$\alpha$ points were discrepant from the remaining points by roughly the equivalent of 1 mag, and/or (4) the questionable points had large uncertainties with respect to the others.  Note that the H$\alpha$ data are prone to being erratic because single \ion{H}{2} regions can drastically affect individual radial bins.

Of the remaining 33 galaxies, 16 had data eliminated in H$\alpha$ as well as in some other band, 11 had either no H$\alpha$ data at all or none by the outer radius of the questionable data, and the remaining six had H$\alpha$ consistent with the general trend at the location of the questionable point in at least one other band.  As with the H$\alpha$ outliers, there were 17 cases where at least one outlier in some band was eliminated because the same clear trend was visible on both sides of the questionable point in that band and frequently in others as well.  There were nine instances (one included in the previous case of 17) where a point or two at either the inside or the outside was eliminated from the fits to prevent an additional break caused by just one slightly high or low point.

Five (of the 33) galaxies (DDO~46, DDO~50, IC~10, IC~1613, and NGC~6822) had outer data in at least two bands eliminated because they appeared to be an additional excess beyond the fitted broken profiles, but not a third exponential component.  DDO~50 alone had 62 data points from 7 passbands (FUV, NUV, $UBV$, H$\alpha$, and 3.6~$\mu$m) eliminated due to 14 annuli with outliers, including 40 points in the excess region in the outermost 7 annuli.  An examination of the 2D images for all five dwarfs indicates that the outer excess is caused by asymmetric SF clumps (as in Figure~\ref{DDO50}) or perhaps even a limited field of view in a few cases.

In the three remaining (of the 33) galaxies (F620-V3, Haro~8, and Mrk~5), each second point from the center was eliminated from all the observed bands because they appeared to be outliers from the general trend.  However, the elimination of these points is particularly important in these three cases because without those outliers, the rest of the data are extremely well fit by a single exponential, \ie they are classified as Type~I.  Because the first point in each of these cases was so identically on the line defined by the third through final points, a broken profile created an awkward overall fit to the data.  Consequently, eliminating the second point in each of these galaxies seems justified because all the remaining data define an excellent single exponential profile.

\begin{figure}
\epsscale{1.1}
\plotone{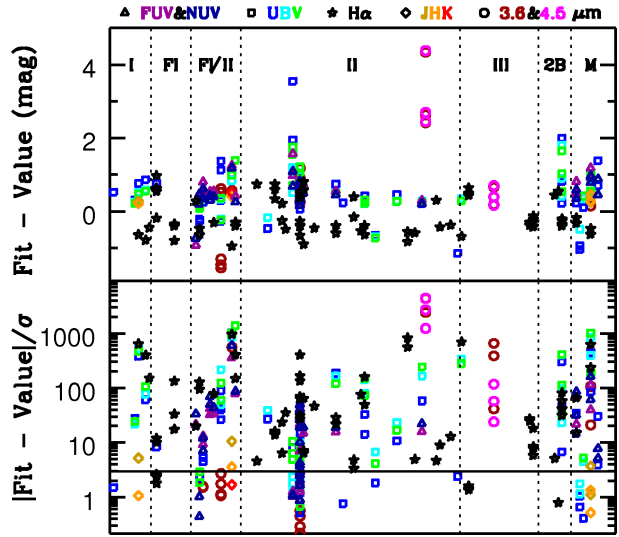}
\caption{Analysis of eliminated data points vs.\,profile type, where the two break profiles and multiple types for different passbands are indicated by 2B and M, respectively.  The upper panel shows the difference between the fit and the data value.  Positive values correspond to excesses (data brighter than the fit) for all bands except for H$\alpha$ where the difference is not in magnitudes but logarithmic variations.  The lower panel shows the absolute value with respect to the uncertainty ($\sigma$) of each data point.  The horizontal line shows 3$\sigma$.  The galaxies are spread out along the $x$-axis for visibility and are grouped by profile type.\label{SigAnal} }
\end{figure}

Figure~\ref{SigAnal} shows the results of an analysis of the discrepancy between the eliminated data points and the final fits.  Table~\ref{TabElim} lists numbers of eliminated and total data points.  In the upper panel of Figure~\ref{SigAnal}, positive values correspond to excesses (where data are brighter than the fit) for all bands, except the reverse is true for H$\alpha$.  All but 20/211 ($\sim$10\%) of the eliminated data points in all the bands except H$\alpha$ were excesses whereas 22/71 ($\sim$31\%) of the eliminated H$\alpha$ data were not excesses.  Of the 20 non-excesses in the non-H$\alpha$ bands, 13 were outer points eliminated to prevent a second break in five galaxies (DDO~70, DDO~154, DDO~216, F533-1, and WLM) and the remaining 7 were low innermost points in three galaxies (DDO~25, DDO~69, and DDO~216).

\begin{deluxetable}{ccccc}
\tabletypesize{\scriptsize}
\tablecaption{Numbers of Eliminated and Total Data Points \label{TabElim}}
\tablewidth{0pt}
\tablehead{\colhead{Band} & \colhead{Excesses\tablenotemark{a}} & \colhead{Eliminated\tablenotemark{b}} & \colhead{Total} & \colhead{Profiles} }
\startdata
FUV	& 28 (11.7\%)		& 29 (4.4\%)	& 655	& 58	\\ 
NUV	& 32 (13.3\%)		& 33 (4.6\%)	& 714	& 61	\\ 
U	& 41 (17.1\%)		& 49 (4.1\%)	& 1209	& 129 \\ 
B	& 27 (11.2\%)		& 32 (2.5\%)	& 1279	& 132 \\ 
V	& 30 (12.5\%)		& 32 (2.3\%)	& 1379	& 141 \\ 
H$\alpha$& 49 (20.4\%)	& 71 (7.6\%)	& 940	& 129 \\ 
J	& 4 (1.7\%)		& 4 (1.3\%)	& 305	& 40	\\ 
H	& 4 (1.7\%)		& 4 (2.5\%)	& 157	& 25	\\ 
K	& 1 (0.4\%)		& 1 (1.2\%)	& 81		& 12	\\ 
3.6~$\mu$m& 18 (7.5\%)	& 21 (4.3\%)	& 492	& 39	\\ 
4.5~$\mu$m& 6 (2.5\%)	& 6 (5.8\%)	& 104	& 10	\\ 
\tableline
all	& 240 (100\%)	& 282 (3.9\%)	& 7315	& 776
\enddata
\tablenotetext{a}{Excess percentages are out of the total number of excesses (240).}
\tablenotetext{b}{Eliminated percentages are out of the total number of points for the relevant band.}
\end{deluxetable}

More data were eliminated in H$\alpha$ than in any other passband.  However, all 141 dwarfs were imaged in $V$ whereas only subsets were imaged in the other 10 filters.  Columns 4 and 5 of Table~\ref{TabElim} show the total number of points and number of profiles per band, respectively.  When the number of eliminated points is normalized by the total number of points per passband, the H$\alpha$ data still stand out as having the most eliminated data points (7.6\%) whereas the FUV, NUV, $U$, and 3.6~$\mu$m data sets have roughly 4-5\% eliminated and then roughly 2-3\% of the $B$ and $V$ data points were eliminated.  The eliminated percentages in the $JHK$ and 4.5~$\mu$m bands are limited by small number statistics, which might also affect the 3.6~$\mu$m numbers.  Regardless, these statistics further support that most of the eliminated data are excesses due to SF clumps because these show up more strongly in H$\alpha$ and the ultraviolet than in redder bands.  Additionally, when data eliminated in multiple bands are taken into account, only six galaxies (DDO~50, DDO~216, DDO~217, DDO~26, UGC~11820, and DDO~87) have eliminated data in more than three annuli.

The lower panel of Figure~\ref{SigAnal} shows the discrepancies with respect to the data uncertainties where the horizontal line indicates 3$\sigma$.  Only 66/282 (23.4\%) of the eliminated data fall below the 3$\sigma$ line.  Of these, only ten had discrepancies less than 0.25 mag and can be broken down into the following: five to prevent a second break in DDO~70 and DDO~169, two eliminated as a general full passband excess in Haro~29, one at the start of the outer excess in DDO~50, and one each in DDO~26 and DDO~210 to allow for better fits. 

\subsubsection{Two Data Point Problem}
As described in Section 3.1, the program splits each profile into every possible pair of continuous subsets containing {\it at least two data points}.  The decision to allow a minimum of {\it two} (instead of three) data points per section was motivated by many visible breaks in {\it multiple} bands after only two data points.  (For example, see the profiles of DDO~155 and DDO~171 in the online version of Figure~1.)  These profiles would have been very difficult to fit without two data segments.  If the first one or two data points had been eliminated in these cases, leaving a Type~I profile, the profile fit would have significantly {\it overestimated} the light at the central most point(s).  More light than the profile suggests can be explained by SF clumps, but less light than the fit has no logical explanation.

Fitting a line to only two data points causes additional complications which are different between the cases of $1/\sigma$ weighting (used for the Type~I fitting and determining the break locations) and no weighting (used for the final broken fits).

{\it $1/\sigma$ Weighting:}  When the minimizing $\chi^2$ function from \citet{NumRec} is used to fit two points with $1/\sigma$ weighting, it uses the weighting to return uncertainties on the slope and $y$-intercept but the goodness-of-fit probability $q$ cannot be calculated because it depends on the number of points subtracted by two.  Also, the default $\chi^2$ is zero.  To prevent a bias in defining breaks with two points on one side over using more data per side, the $\chi^2$ values were changed for the two points case.  In general, $\chi^2$ is increased by $(y - \rm{fit})^2/\sigma$.  For two points, we assume that $y - \rm{fit} \sim \sigma$ so $\chi^2 \simeq \sigma_1 + \sigma_2$.  This adjustment appears to give reasonable values of $\chi^2$ to counteract the bias toward breaks with two points on one side.

{\it No Weighting:}  The $q$ and $\chi^2$ values are again problematic as well as the uncertainties on the slope and $y$-intercept.  The default function returns 1.0 for $q$ in the no weighting case, but for our purposes we changed this value to 2.0 as a visual tag for the two point case.  Here $\chi^2$ is increased by $(y - \rm{fit})^2$.  If we again assume that $y - \rm{fit} \sim \sigma$ for the two point case, then $\chi^2 \simeq \sigma_1^2 + \sigma_2^2$.

No weighting yields default slope and $y$-intercept uncertainties of zero.  Since the data have $y$ uncertainties, consider two extreme cases of fitting a line between (1) $(x_1, y_1-\sigma_1)$ and $(x_2, y_2+\sigma_2)$ to get slope $m_1$ and $y$-intercept $b_1$ as well as (2) $(x_1, y_1+\sigma_1)$ and $(x_2, y_2-\sigma_2)$ to get slope $m_2$ and $y$-intercept $b_2$.  The ``best'' fit line will go through the two points, resulting in a slope and $y$-intercept that are between $m_1$ and $m_2$ and $b_1$ and $b_2$, respectively.  While there is no reason to assume that the ``best'' values are {\it half-way} between the two extremes, a reasonable uncertainty for the slope and $y$-intercept values is half the difference between the two extreme values, \ie $|m_1 - m_2|/2$ and $|b_1 - b_2|/2$, respectively.

Very few outer sections were fit with only two points: 0 for $JK$, 3.6 and 4.5~$\mu$m, 1-2\% for FUV, NUV, and $UBV$, 4\% for $H$ (small number statistics) and 15\% for H$\alpha$.  The relatively large percentage in H$\alpha$ is due to the limited extent of the H$\alpha$ profiles.  Significantly more inner sections were fit with only two points: 0 for $K$, 7-14\% for FUV, NUV, $UBVJH$ and 3.6~$\mu$m, 20\% for 4.5~$\mu$m, and 29\% for H$\alpha$.  The 4.5~$\mu$m statistics are severely limited by the small sample and the H$\alpha$ statistics reflect the dominance of resolved sources in the azimuthal averaging.

\subsubsection{Fitting Two Breaks to Profiles from Nine Galaxies}
Our final complication is that we found nine galaxies in our sample of 141 (6\%) whose profiles are better fit by three lines instead of two, and consequently contain two breaks.  Doubly broken profiles, though relatively uncommon, are not unheard of.  In two large samples of spiral galaxies, a handful of so-called {\it mixed} profiles have been identified.  Out of a sample of 85 face-on to intermediately inclined late-type (Sb - Sdm) spirals, seven (8\%) were discovered to be Type II then III and two (2\%) were classified as Type II then II again (PT06).  Furthermore, four galaxies (6\%) and five galaxies (11\%) were classified as Type II then III in a sample of 66 barred early-type (S0 - Sb) galaxies (EPB08) and 47 unbarred early-type (S0 - Sb) galaxies (GEAB11), respectively.  Interestingly, none of these studies found a spiral Type~III galaxy with an additional outer break.

Eight of the nine two break galaxies in our study first stood out when the program fit one break in some bands but a different break in others.  Visibly both breaks were apparent in more than one band.  The ninth galaxy, DDO~125, only has data in $V$ and the more erratic H$\alpha$.  However, the one break fit to the DDO~125 $V$ band data was visibly poorer than most of the one break fits whereas the two break fit cleanly goes through every point.

The basic program was not sophisticated enough to fit two breaks, so we forced it to fit three pieces by giving the data in two overlapping sections.  In this manner, the program determined the location of the first break from the inner and middle data and then the second break from the middle and outer data.  Since it fit the middle section twice, the scale length and $y$-intercept uncertainties from the two versions of the fit were compared for each passband that contained two breaks, a total of 39 profiles.  Initially some differences were found due to the somewhat arbitrary decision of how to split the data for the two breaks.  However, in almost every case, the fit with the lower uncertainty on the scale length also had the lower uncertainty on the $y$-intercept.  This trend was applied to make the program use the better middle fit.  The input was also iteratively adjusted to improve agreement between the two fits of the middle sections.

\subsubsection{Classification of Broken Profiles: II, III, and FI}
Once all the fitting was done, the next task was to subdivide the broken profiles into Types~II (down-bending), III (up-bending), and FI (flat inside or increasing).  The most difficult subdivision was isolating the FI profiles from their parent II type.  Empirically the FIs were assigned to profiles having an inner slope less than 0.075, corresponding to an inner scale length that is either greater than $\sim$14.5 kpc or negative due to the surface brightness increasing from the center to the break.  Virtually all the profiles where the uncertainty on the inner scale length was larger than the value itself also appeared relatively flat inside, so this was an additional qualification for Type~FI.  The subdivision of the remaining broken profiles as either Type~II or III was determined by a simple comparison between the inner and outer scale lengths (or slopes) such that IIs are steeper beyond the break and thus have a shorter outer scale length (larger slope) and IIIs are the reverse.  (See Table \ref{tabII_III} for some defining relationships about Type~II and III profiles.)

\begin{deluxetable}{ccc}
\tabletypesize{\scriptsize}
\tablecaption{Defining Relationships \\ for Types II and III\label{tabII_III}}
\tablewidth{0pt}
\tablehead{\colhead{Characteristic} & \colhead{II} & \colhead{III} }
\startdata
outside		& steeper		& shallower	\\
slope		& in $<$ out	& in $>$ out	\\
scale length	& in $>$ out	& in $<$ out	\\
$\mu_{0,o}$ wrt $\mu_{0,i}$	& brighter		& fainter	\\
$\mu_0$   	& in $>$ out	& in $<$ out
\enddata
\end{deluxetable}

All the measurements are in magnitudes that {\it increase} with radius {\it except} for the $\log \Sigma$ data for H$\alpha$.  Since these values {\it decrease} with radius, the sign conventions for the profile classification are inverted.  The H$\alpha$ profile type was frequently disregarded when determining the overall profile type of each galaxy, listed in Column 8 of Table~1.  In some instances, some bands had one profile type while at least one other had a different type.  These galaxies, with {\it multiple} profile types, are considered separately throughout the analysis.  Since FI is just a special case of Type~II profiles, galaxies with a combination of FI and II are not considered to have multiple profile types.  However, since there are enough FI/II galaxies, they are considered a subcategory of Type II profiles.

\subsection{Comparison to Previous Fits}
\begin{figure}
\epsscale{0.85}
\plotone{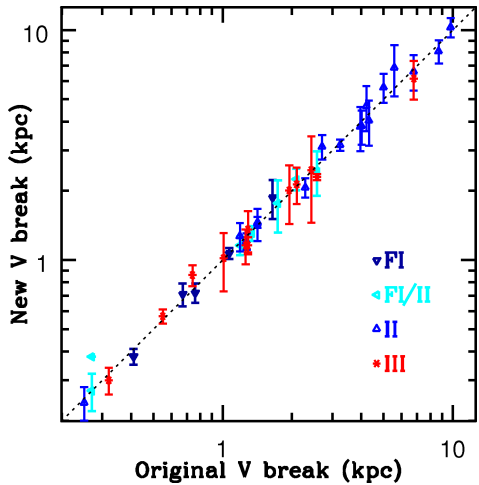} 
\caption{Comparison between the original $V$-band break locations determined by Hunter \& Elmegreen (2006) and the values determined here.  The original break radii have been corrected to the distances used here (see Table 1).  Virtually all the points are equal within the uncertainties on the new values.  The only clear outlier (a FI/II data point at the lower left) corresponds to the first of two breaks in Mrk~178 whereas only one break was originally used.  \label{OrigNew} }
\end{figure}

One obvious method of checking the results of the fitting is to compare the breaks to those determined previously in Hunter \& Elmegreen (2006).  Although fewer profiles were fit with breaks in that study, there is still a large enough sample for comparison.  Figure~\ref{OrigNew} shows the original $V$-band break locations plotted as a function of the values determined here.    Within the uncertainties, there is excellent agreement.  The only clearly outlying data point (a FI/II break in the lower left) belongs to Mrk~178.  The profile of this BCD was originally fit with one break whereas it is one of the few two break profiles in this study.  Consequently, the discrepancy is not surprising.  Several other points are only just barely discrepant beyond their uncertainties.

\begin{figure*}
\epsscale{0.9}
\plotone{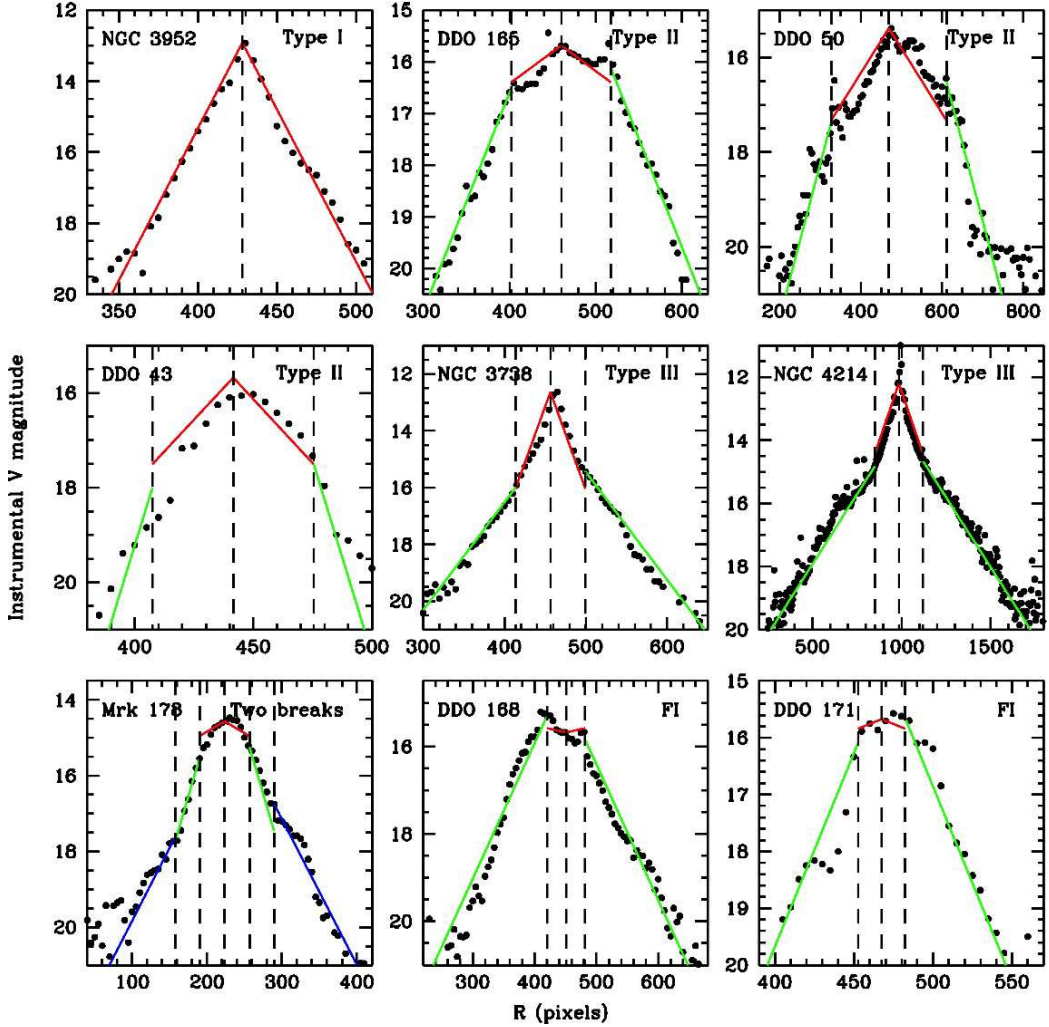} 
\caption{Cuts along the major axis of a few representative galaxies, in instrumental $V$ magnitudes (filled circles).  The azimuthally-averaged surface photometry fits are superposed; {\it red} is the inner fit; {\it green}, the outer; and {\it blue} the outermost fit for the galaxy with two breaks. The vertical dashed lines mark the morphological center and the breaks at the appropriate radius on either side of center.  The lines have been anchored to the central points, but are sometimes slid up or down at the breaks.  Note that the breaks appear on both sides of each broken galaxy even though these dwarfs are irregular.  \label{Cuts} }
\end{figure*}

\subsection{Asymmetries?}
These are all irregular galaxies, so an important aspect to consider is if our first analysis step, azimuthal averaging, is even justified.  To some extent, the intriguing trends (or consistencies) with $M_B$, profile type, and wavelength in Section 4 indicate that the fit parameters (central and break brightnesses, scale lengths, and break locations) are physically meaningful.  (Complete scatter plots could have indicated some potentially serious problems!)  However, since the ends do not necessarily justify the means, we explored the nature of the breaks from simple cuts along the major axis of a few representative galaxies in Figure~\ref{Cuts}.  From these cuts, we can examine whether asymmetries in the galaxy might be creating the apparent breaks in the surface photometry profiles.

\begin{deluxetable*}{cccccccccccccccc}
\tabletypesize{\scriptsize}
\tablecaption{Fit Parameters\label{FitParams}}
\tablewidth{0pt}
\tablehead{ \colhead{Galaxy} & \colhead{Band} & \colhead{Type}
& \colhead{$h_{R,i}$} & \colhead{$\Delta h_{R,i}$} & \colhead{$\mu_{0,i}$} & \colhead{$\Delta\mu_{0,i}$}
& \colhead{$R_{br}$} & \colhead{$\Delta R_{br}$} & \colhead{$\mu_{br}$} & \colhead{$\Delta\mu_{br}$}
& \colhead{$h_{R,o}$} & \colhead{$\Delta h_{R,o}$} & \colhead{$\mu_{0,o}$} & \colhead{$\Delta\mu_{0,o}$}
& \colhead{Notes} \\
& & & a,b & a & c & c & a & a & c & c & a,b & a & c & c & d
}
\startdata
0467-074 & U & II & 4.88 & 2.39 & 22.97 & 0.41 & 6.64 & 5.48 & 24.44 & 1.48 & 2.18 & 0.53 & 21.13 & 0.97 & \nodata \\
\nodata & B & II & 5.36 & 0.39 & 23.20 & 0.06 & 5.96 & 2.43 & 24.40 & 0.50 & 2.59 & 0.32 & 21.91 & 0.41 & \nodata \\
\nodata & V & II & 4.35 & 0.38 & 22.63 & 0.08 & 6.13 & 3.12 & 24.15 & 0.79 & 2.72 & 0.30 & 21.71 & 0.35 & \nodata \\
\nodata & Ha & FI & 84.95 & 539.53 & 32.04 & 0.31 & 6.09 & 3.66 & 32.12 & 0.58 & -7.16 & 0.72 & 33.04 & 0.12 & \nodata \\
1397-049 & B & II & 5.32 & 0.52 & 23.96 & 0.06 & 4.48 & 0.46 & 24.88 & 0.14 & 2.43 & 0.02 & 22.88 & 0.02 & \nodata \\
\nodata & V & II & 4.40 & 0.21 & 23.27 & 0.03 & 4.47 & 1.04 & 24.38 & 0.26 & 2.10 & 0.14 & 22.06 & 0.22 & \nodata \\
A1004+10 & B & III & 0.33 & 0.02 & 20.70 & 0.17 & 1.72 & 0.54 & 26.29 & 1.79 & 0.45 & 0.02 & 22.16 & 0.19 & 1;ie \\
\nodata & V & III & 0.35 & 0.01 & 20.46 & 0.10 & 1.70 & 0.21 & 25.76 & 0.69 & 0.57 & 0.02 & 22.52 & 0.14 & 1;ie \\
\nodata & Ha & II & -1.52 & 0.20 & 33.16 & 0.07 & 1.00 & 1.08 & 32.45 & 0.78 & -0.45 & 0.23 & 34.87 & 1.36 & 1;*ie;2o \\
A2228+33 & U & III & 1.31 & 0.00 & 22.33 & 0.00 & 2.22 & 0.46 & 24.17 & 0.38 & 2.53 & 0.10 & 23.22 & 0.17 & \nodata \\
\nodata & U & II & \nodata & \nodata & \nodata & \nodata & 7.87 & 1.56 & 26.59 & 0.73 & 1.12 & 0.07 & 18.98 & 0.60 & \nodata \\
\nodata & B & III & 1.31 & 0.00 & 22.39 & 0.00 & 2.57 & 0.27 & 24.52 & 0.22 & 2.89 & 0.08 & 23.55 & 0.11 & \nodata \\
\nodata & B & II & \nodata & \nodata & \nodata & \nodata & 7.04 & 2.43 & 26.20 & 0.93 & 1.18 & 0.15 & 19.72 & 1.03 & \nodata \\
\nodata & V & III & 1.34 & 0.00 & 21.91 & 0.00 & 2.69 & 0.21 & 24.09 & 0.17 & 2.91 & 0.06 & 23.09 & 0.08 & \nodata \\
\nodata & V & II & \nodata & \nodata & \nodata & \nodata & 6.93 & 2.47 & 25.67 & 0.93 & 1.43 & 0.16 & 20.41 & 0.75 & \nodata \\
\nodata & Ha & III & -2.34 & 0.83 & 32.96 & 0.14 & 1.11 & 1.01 & 32.45 & 0.52 & -5.35 & 0.31 & 32.67 & 0.12 & 2i \\
\nodata & Ha & III & \nodata & \nodata & \nodata & \nodata & 12.82 & 98.74 & 30.07 & 20.02 & -5.10 & 1.53 & 32.80 & 0.39 & \nodata
\enddata
\tablenotetext{}{Notes: (1) Table~\ref{FitParams} is available in its entirety in the electronic edition of the {\it Astronomical Journal}.  A portion is shown here for guidance regarding its form and content.  (2) A profile of Type~I has no break location nor outer fit, so columns 8-15 contain no data.  (3) A profile that is fit with two breaks is listed on two lines such that the first line contains the inner fit parameters, the information for the first break, and the middle fit parameters while the second line contains only the information about the second break and the outermost fit parameters.  Because the middle fit parameters do not need to be given twice, columns 4-7 contain no data in the second line for profiles with two breaks.}
\tablenotetext{a}{In kpc.}
\tablenotetext{b}{A {\it negative} scale length in any {\it broadband} filter indicates a scale length such that the light is {\it brightening}; these are only found in the inner scale lengths, $h_{R,i}$.  Alternately, a {\it negative/positive} scale length in H$\alpha$ indicates a scale length such that the light is {\it dimming/brightening}; can be found in either the inner or outer scale lengths, $h_{R,i}$ or $h_{R,o}$.}
\tablenotetext{c}{In mag~arcsec$^{-2}$, except for H$\alpha$ data which are $\log \Sigma_{\rm{H} \alpha}$ in erg~s$^{-1}$~pc$^{-2}$.}
\tablenotetext{d}{Notes column: \\
- ``\#;'' = the number of eliminated points in that band's profile; \\
- a series of the following letters possibly preceded by an * explain the eliminated data such that: \\
- i/m/o = eliminated data point(s) in the inner/middle/outer third of the profile, \\
- e/d = eliminated data point(s) are an e=excess (brighter than the fit) or d=decrement (fainter than the fit), \\
- ``*'' = a significant discrepancy from the fit; \\
- B = profile forced to be broken whereas S = profile forced to single; and \\
- an i, m, or o (or a combination thereof) preceded by a 2 indicates the inner, middle, or outer segment (or a combination) were fit by only two data points. }
\end{deluxetable*}

We chose a few dwarfs that exhibit the major profile types: I, II, III, FI (flat or increasing inside), and those with two breaks.  We rotated a subsample of $V$-band images about the galaxy center to put the major axis along a row.  From each rotated image we extracted a 1D cross cut along the major axis, summing over an integer number of rows close to 10\arcsec\ in width, in order to increase the signal-to-noise ratio (S/N).  The counts in the cut were converted to instrumental magnitudes.  We then averaged every five points along the row, again to increase S/N.  These cuts are plotted in Figure~\ref{Cuts} as the filled circles.  The superposed solid lines are the surface photometry fits.  We anchored each line segment approximately to the data points at the top of the line.  Some line segments were slid up or down to better match the data.  The morphological center of the galaxy and the break radii are marked by the vertical dashed lines.

We found that the breaks appear on both sides of each galaxy.  DDO~43 is included as an example where things are messier than usual, probably due to low S/N, but even there the breaks can be seen on both sides of center.  Additionally, we have visually checked virtually all the 2D images in $V$ with the elliptical bins used in the azimuthal averaging step superimposed.  (See Figure~\ref{DDO50} for an example.)  Although these are irregular galaxies, almost all of them have at least outer profiles that appear extremely elliptical with well-determined centers, principal axes, and $b/a$ values.  Consequently it seems that the azimuthally averaged radial profiles truly reveal how the light is generally falling off with radius, especially since the outlying annuli from overly bright SF clumps have generally been eliminated from the analysis.  We will examine possible relationships between profile types and the 2D images in a future paper, including additional attention to asymmetries as well as the possible importance of SF clumps to the FI profile type.

\section{RESULTS}
Figure~\ref{Rep7} displays the surface brightness profiles with their fits, as well as the radial colors, of a representative dwarf of each different profile type: I, II, FI, FI/II, III, multiple types for various bands, and two breaks (frequently referred to as {\it mixed} in the literature).  Azimuthally-averaged surface brightness plots of the full sample of 141 dwarfs are available in the online version.  Table~\ref{FitParams} lists all the fit parameters for the profiles for each observed passband: Type, inner scale length $h_{R,i}$, central surface brightness $\mu_{0,i}$, break location $R_{br}$, surface brightness at the break $\mu_{br}$, outer scale length $h_{R,o}$, and outer surface brightness projected back to the center $\mu_{0,o}$, along with an uncertainty for each parameter, and some notes, including the number of eliminated points.  The parameters for the outer most fit in the nine galaxies with two breaks are listed in a second row per passband.  Profiles fit by a single line logically contain no break or outer fit information.

\subsection{Fit Parameters vs.\,$M_B$, Profile Type, and Passband}
In general, each broken profile is fit by four parameters: the $y$-intercept and slope for two lines, where the disk scale length is $h_R = 2.5\log(e)/$slope.  The $y$-intercept values are the surface brightnesses from projecting the inner ($\mu_{0,i}$) and outer ($\mu_{0,o}$) fits back to radius~=~0.  The break radius, $R_{br}$, and the break surface brightness, $\mu_{br}$, are also of interest.  Type~I profiles are fit by a single line, so they do not have a break or outer profile.  However, it is interesting to compare the $\mu_0$ and $h_R$ values of the Is to both the $\mu_{0,i}$ and $h_{R,i}$ values and the $\mu_{0,o}$ and $h_{R,o}$ values from the broken profiles, where the $i$ and $o$ subscripts indicate the inner and outer fits, respectively.  Also, the nine galaxies with two breaks have information for the middle section such that the outer parameters ($\mu_{0,o}$ and $h_{R,o}$) for the first break are equal to the inner parameters ($\mu_{0,i}$ and $h_{R,i}$) for the second break.  In the following figures, ``2Bi'' and ``2Bo'' points indicate data for the first (inner) and second (outer) breaks, respectively.

Below, we discuss the above parameters as a function of $M_B$, profile type, and passband.  For the three surface brightness values ($\mu_{0,i}$, $\mu_{0,o}$, and $\mu_{br}$), the wavelength dependent discussions are considered as colors with respect to the relevant values in $V$: the colors obtained from projecting the inner and outer profiles back to radius~=~0 and the color at the break.  Radial color trends will be the focus of another paper in this series.

Note that 9/141 dwarfs in this study are missing $M_B$ values (no $B$ data) and thus are not in the following plots.  These nine dwarfs were sacrificed to plot information about late-type spirals (from PT06) and early-type spirals (from EPB08 and GEAB11), all with $M_B$ data, to examine trends between dwarfs and spirals.  The original data on the 85 late-type spirals in PT06 are in $g'$ and $r'$ filters, so $V = g'-0.55(g'-r')-0.03$ \citep{SDSSfilters} was used to transform the surface brightness data to $V$ and the $g'$ scale length and break location data are compared to the dwarf $V$ results.

Since the numbers of galaxies with $HK$ and 4.5~$\mu$m data are relatively small (25, 12, and 10, respectively), none of the following plots show results from these passbands.  To explore a sampling of passbands including H$\alpha$, important to SF, plots are shown for FUV, $V$, 3.6~$\mu$m, and H$\alpha$ when single passband data are plotted and differences of FUV, $U$, $3.6~\mu$m, and either $J$ or H$\alpha$ with respect to $V$ to examine dependences on wavelength.  Please keep in mind that at least some of the scatter in the plots could be caused by uncertainties in the distances to these galaxies, affecting $M_B$.

\begin{figure*}
\epsscale{0.98}
\plotone{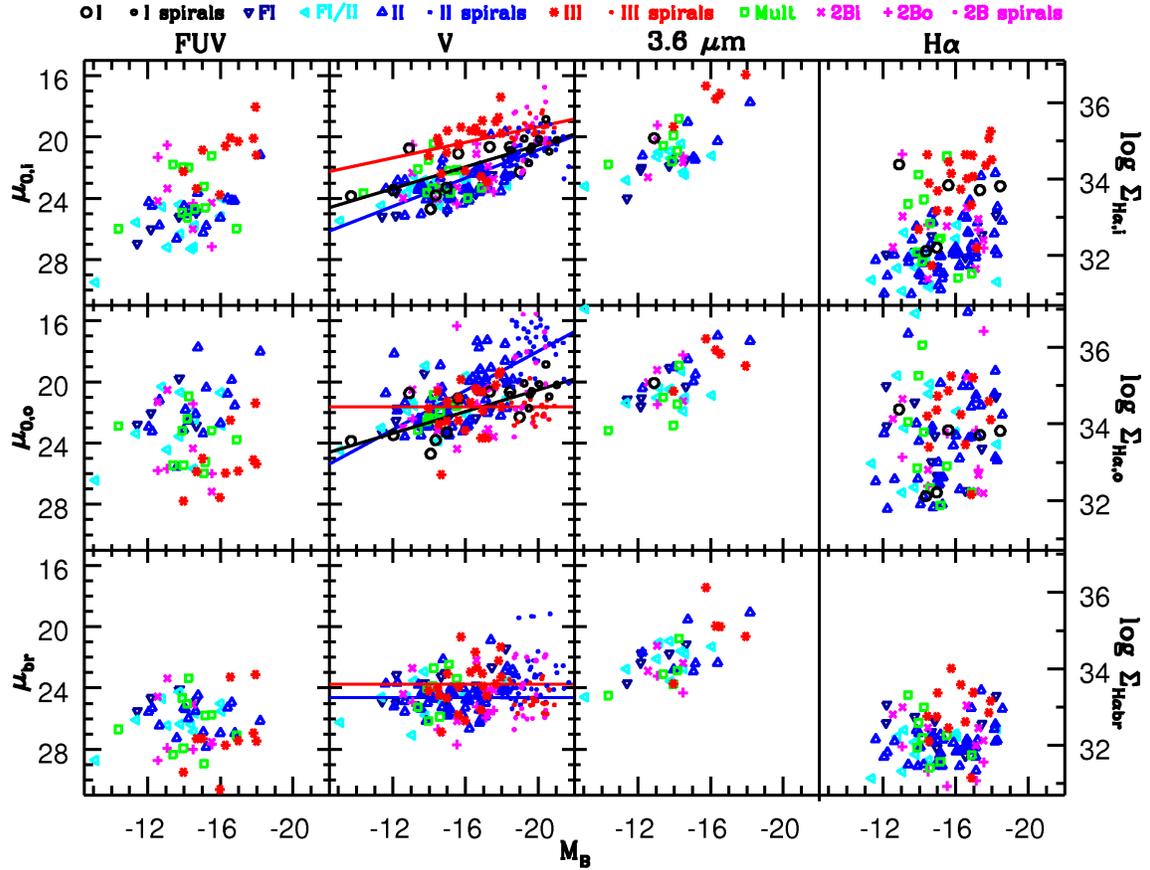}
\caption{Three surface brightnesses: {\it (top)} projected central value from the inner exponential, $\mu_{0,i}$, {\it (middle)} projected central value from the outer exponential, $\mu_{0,o}$, and {\it (bottom)} value at the break location, $\mu_{br}$.  In the columns, three representative passbands are shown as well as H$\alpha$.  Note that the first three columns are in mag~arcsec$^{-2}$ whereas the H$\alpha$ data are logarithmic surface brightnesses in erg~s$^{-1}$~pc$^{-2}$.  Since Type~I profiles have no second component, their central surface brightnesses are plotted in both the top and middle panels for comparison.  The $V$-band data for the spirals are transformed from $g'$ and $r'$ in PT06 as indicated in the text.  Note the clear trends with $M_B$ (and the lack thereof in $\mu_{br}$), the separation between Types~II and III (especially between dwarfs in the top row versus spirals in the middle), and comparisons between dwarfs and spirals.  See Table~\ref{mu_trends} for details about the linear fits in the $V$ panels.  The sloped lines show the least-square fits of the dwarf and spiral samples for Type~I, II, and III, in the same color as the corresponding data.  The one blue and two red horizontal lines in the $V$ bottom and middle panels show the dwarf averages.  In the legend, ``Mult'' and ``2B'' signify galaxies fit by multiple profile types and two breaks, respectively, and ``2Bi'' and ``2Bo'' indicate the inner and outer break parameters, respectively. \label{Mu} }
\end{figure*}

\subsubsection{Central Surface Brightness: $\mu_{0,i}$}
The top row of Figure~\ref{Mu} shows central surface brightness values extrapolated from the inner fits in FUV, $V$, 3.6~$\mu$m, and H$\alpha$ as functions of $M_B$.  Clearly more luminous dwarfs have brighter centers and dwarfs are brighter in redder bands, even though dwarfs are generally bluer than spirals.  However, Type~III galaxies have brighter centers than Is or IIs but all three types follow roughly parallel trends with $M_B$.  Some of the Type~I dwarfs are more similar to the IIIs whereas others agree better with the Type~IIs.  Similar trends are seen in NUV and in $UBJ$.  By coincidence, we are missing ultraviolet data for every galaxy with a pure Type~I profile.

The small points at the bright end of the $V$ panel are PT06 spirals transformed from $g'$ and $r'$.  The lines in the $V$ panels show least-square fits to the dwarf and spiral Type~I, II, and III samples, colored the same as the data.  Table~\ref{mu_trends} gives the line parameters.  All the spirals (I, II, and III) follow the same trends with $M_B$ as their dwarf counterparts.  The two dwarf branches of Types~II and III are more separated than the spirals, but the spiral IIIs are still on the bright end with the Is falling in between.  A K-S 1D two sample test gives a probability of $5.8\times10^{-8}$ that the dwarf Types~II and III follow the same distribution and a probability of $3.7\times10^{-5}$ for the spirals, so the Types~II and III really are distinct.

\subsubsection{Outer Projected Central Surface Brightness: $\mu_{0,o}$}
The middle row of Figure~\ref{Mu} shows the outer projected central surface brightness values for the same four passbands, again as a function of $M_B$, and with the same vertical axis.  More luminous Type II dwarfs have brighter $\mu_{0,o}$ values than less luminous dwarf IIs.  Unlike $\mu_{0,i}$, the Type~III $\mu_{0,o}$ values are fainter than Type~IIs for bluer (i.e., FUV) bands and gradually become more similar to the other types at redder bands.  (This trend is more apparent when the NUV, $U$, and $B$ results are also examined.)  In $V$, the dwarf Types~II and III generally overlap (with a high 1D K-S same distribution probability of 37\%) and agree with the $\mu_0$ values of the Type~Is, with K-S probabilities of 1.5\% and 15\% that the Types~II and III follow the same distribution as the Is, respectively.

Again the comparison between dwarfs and spirals in $V$ yields interesting results.  As with $\mu_{0,i}$, the Type~II spirals continue the brightening $\mu_{0,o}$ trend of the Type~II dwarfs whereas for the Type~IIIs, $\mu_{0,o}$ is approximately 21.7 mag~arcsec$^{-2}$ for both dwarfs and spirals (see Table~\ref{mu_trends}).  Here the {\it spirals} have more separated trends (low K-S same distribution probability of $5.9\times10^{-13}$) between the Types~II and III, with the spiral Is in between.

\begin{deluxetable}{cccccc}
\tabletypesize{\scriptsize}
\tablecaption{Linear Trend Fits of Surface \\ 
Brightness Parameters in $V$\label{mu_trends}}
\tablewidth{0pt}
\tablehead{ \colhead{Parameter$^a$} & \colhead{Type} & \colhead{Slope$^b$} & \colhead{y-int$^c$} & \colhead{Scatter$^d$} & \colhead{} }
\startdata
$\mu_0$     &		I    &	0.354 &27.6 & 1.07 & \\
$\mu_{0,i}$ &	II$^e$   &	0.459 &30.0 & 0.96 & \\
$\mu_{0,i}$ &	III  	     &	0.252 &24.4 & 1.01 & \\
$\mu_{0,o}$ &	II$^e$   &	0.640 &30.8 & 1.52 & \\
\tableline
Parameter$^f$ & Type & Dwarf & Dwarf & Spiral & Spiral \\
& & Ave & $\sigma$ & Ave &  $\sigma$\\
\tableline
$\mu_{0,o}$&	III  &	21.6 & 0.4 & 21.8 & 0.2 \\
$\mu_{br}$ &	II$^e$   &	24.6 & 0.1 & 23.3 & 0.2 \\
$\mu_{br}$ &	III  &	23.7 & 0.3 & 24.7 & 0.2
\enddata
\tablenotetext{$^a$}{These trends were determined from a combination of the dwarf and spiral data.}
\tablenotetext{$^b$}{The slope is in $V$-band magnitudes per $M_B$ magnitudes.}
\tablenotetext{$^c$}{The $y$-intercept is in $V$-band magnitudes.}
\tablenotetext{$^d$}{The scatter is the standard deviation in the data when the trend is subtracted out.}
\tablenotetext{$^e$}{These statistics are for the sample of {\it pure} Type~IIs only.  The dwarf FI and FI/II (as well as two break and multiple profile) type samples have been ignored.}
\tablenotetext{$^f$}{These data were relatively independent of $M_B$, so averages and uncertainties in the mean are given instead of trend fits.}
\end{deluxetable}

\subsubsection{Surface Brightness at Breaks: $\mu_{br}$}
The surface brightness at the break is shown in the bottom row of Figure~\ref{Mu}.  As with $\mu_{0,o}$, the Type~III break surface brightness values are fainter than Type~IIs for bluer (i.e., FUV) bands and gradually become closer to the other types at redder bands.  In the FUV, the Type~IIIs may have a steep brightening relationship with $M_B$, but the IIs have a slightly inverted relationship such that more luminous disks are {\it fainter} at the break location.

\begin{figure*}
\plotone{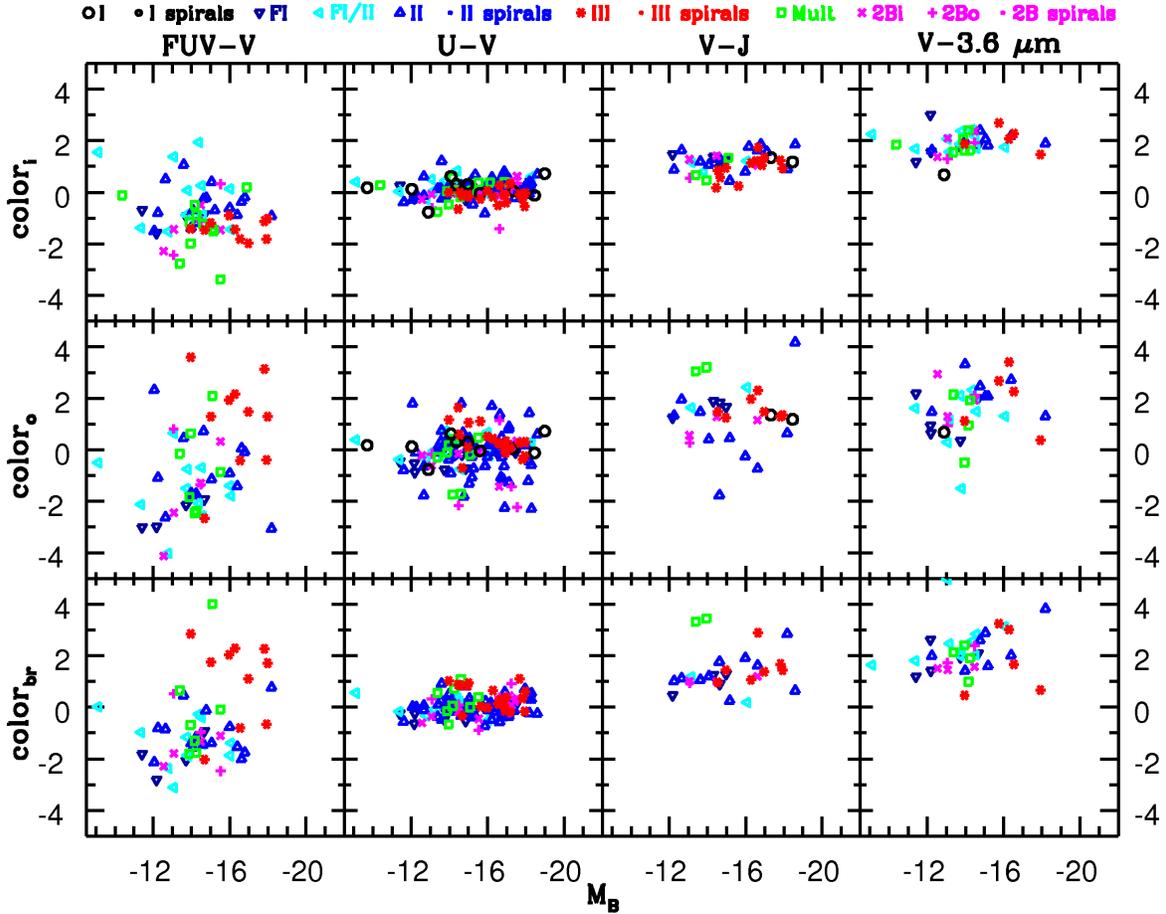}
\caption{Four representative colors at three specific locations: {\it (top)} projected central value from the inner fit, color$_i$, {\it (middle)} projected central value from the outer fit, color$_o$, and {\it (bottom)} value at the break, color$_{br}$.  Since the FUV data are in AB mags, a $-2.46$ correction (from http://mips.as.arizona.edu/$\sim$cnaw/sun.html) has been applied to shift the data to the Vega scale.  As with Figure~\ref{Mu}, the inner values for the Type~I galaxies are also plotted in the middle row for comparison.  Note that the colors are relatively independent of $M_B$ and average values are given in Table~\ref{tabAveCol}. \label{Color} }
\end{figure*}

\begin{deluxetable*}{lcccccccccccccccccc}
\tabletypesize{\scriptsize}
\tablecaption{Average Dwarf Colors \label{tabAveCol}}
\tablewidth{0pt}
\tablehead{
 & \multicolumn{3}{|c|}{FUV-V$^a$} & \multicolumn{3}{|c|}{NUV-V$^a$} & \multicolumn{3}{|c|}{U-V} & \multicolumn{3}{|c|}{B-V} & \multicolumn{3}{|c|}{V-J} & \multicolumn{3}{|c|}{V-3.6~$\mu$m} \\
\colhead{Profile} & \colhead{\#} & \colhead{Ave} & \colhead{$\sigma^b$} & \colhead{\#} & \colhead{Ave} & \colhead{$\sigma^b$} & \colhead{\#} & \colhead{Ave} & \colhead{$\sigma^b$} & \colhead{\#} & \colhead{Ave} & \colhead{$\sigma^b$} & \colhead{\#} & \colhead{Ave} & \colhead{$\sigma^b$} & \colhead{\#} & \colhead{Ave} & \colhead{$\sigma^b$} }
\startdata
I: in    & 0  & \dots & \dots & 0  & \dots & \dots & 10 & 0.14  & 0.13 & 10 & 0.41 & 0.04 & 2  & 1.27 & 0.09 & 1 & 0.68  & \dots \\
II$^c$: in   & 17 & -0.36 & 0.21 & 19 & 0.03 & 0.21 & 56 & 0.19  & 0.05 & 57 & 0.41 & 0.02 & 13 & 1.40 & 0.14 & 8 & 1.84  & 0.18 \\
III: in  & 10 & -1.42 & 0.11 & 10 & -0.94 & 0.11 & 19 & -0.15 & 0.06 & 21 & 0.26 & 0.02 & 12 & 0.95 & 0.13 & 5 & 2.08  & 0.20 \\
II$^c$: out  & 16 & -0.93 & 0.51 & 18 & -0.44 & 0.51 & 55 & -0.14 & 0.12 & 56 & 0.48 & 0.09 & 10 & 1.01 & 0.52 & 8 & 1.91  & 0.37 \\
III: out & 10 & 1.14 & 0.59 & 10 & 0.55 & 0.59 & 19 & 0.36  & 0.13 & 21 & 0.50 & 0.12 & 7  & 1.58 & 0.15 & 5 & 1.97  & 0.54 \\
II$^c$: br  & 16 & -0.89 & 0.26 & 18 & -0.54 & 0.26 & 55 & -0.01 & 0.05 & 56 & 0.28 & 0.03 & 10 & 1.30 & 0.22 & 8 & 2.16  & 0.33 \\
III: br & 10 & 1.06 & 0.52 & 10 & 0.80 & 0.52 & 19 & 0.39  & 0.11 & 21 & 0.50 & 0.08 & 7  & 1.55 & 0.24 & 5 & 1.82  & 0.57
\enddata
\tablenotetext{$^a$}{The FUV and NUV data have been adjusted by -2.46 and -1.86, respectively (from http://mips.as.arizona.edu/$\sim$cnaw/sun.html), to shift from AB magnitudes to the Vega system.}
\tablenotetext{$^b$}{These are the uncertainties in the mean.}
\tablenotetext{$^c$}{These statistics are for the sample of {\it pure} Type~IIs only.}
\end{deluxetable*}

In $V$, the break surface brightness is $\sim$24 mag~arcsec$^{-2}$ for dwarfs and the PT06 spirals of all profile types, though with significant scatter (see Table~\ref{mu_trends}).  The Type~II galaxies appear to occur in two clumps: galaxies with $|M_B| \simlt 17.5$ have $V$-band $\mu_{br}$ values between $26-23.5$ mag~arcsec$^{-2}$ whereas galaxies with $|M_B| \simgt 17.5$ have values between $25-22$ mag~arcsec$^{-2}$, with a few bright outliers.  Since only 6/86 Type~II dwarfs with $\mu_{br} < 23$ mag~arcsec$^{-2}$ occur in this second clump, the apparent pattern may be caused by small number statistics.  As with $\mu_{0,o}$, the dwarf Types~II and III are relatively similar (with a K-S same distribution probability of 0.28\%) but the Type~IIIs are slightly brighter than the IIs.  The $\mu_{br}$ values for the PT06 spirals are less similar (with a K-S same distribution probability of 0.011\%) but with the Type~IIIs being fainter than the IIs, though with significantly less separation than in the case of $\mu_{0,o}$.

\subsubsection{Central Colors}
The top row of Figure~\ref{Color} shows four central colors, extrapolated from the inner fits, as functions of $M_B$.  For each color, the values are relatively independent of $M_B$ though with some scatter.  Table~\ref{tabAveCol} lists the average values for the colors shown as well as for $NUV-V$ and $B-V$.  The difference between the color panels in Figure~\ref{Color} is consistent with the redder bands being brighter than the bluer bands on the Vega scale.  (Note the shift from a bluer band subtracted by $V$ in the left two columns to $V$ minus a redder band in the right two columns.)  The dwarf Type~IIIs are generally bluer than the other types at all colors except $V-3.6~\mu$m.  This trend is related to the radial color profiles that will be explored elsewhere.

No spiral data were available for these plots since the PT06 survey only used $g'$ and $r'$ filters.

\subsubsection{Outer Projected Central Colors}
The middle row of Figure~\ref{Color} shows the same four colors as functions of $M_B$ but determined from the outer projected surface brightness values.  These colors are not as apparently meaningful unless some outer component is representative of an underlying stellar population.  The scatter is greater than with the inner projected central colors and here the Type~III dwarfs generally overlap or are redder than the other types at all passbands.  However, the values are still roughly independent of $M_B$.  See Table~\ref{tabAveCol} for the averages.

\subsubsection{Break Colors}
The same four colors, but at the profile break, are shown in the bottom row of Figure~\ref{Color}, again as a function of $M_B$.  As with the central colors, the break colors are relatively constant with $M_B$ (see Table~\ref{tabAveCol}) except for a slight trend that more luminous galaxies have slightly redder colors.  The scatter is again large.  Unlike the central colors (color$_i$), at the break the Type~III dwarfs are generally redder than the other types at all colors except $V-3.6~\mu$m.  Again, this trend is related to radial color profiles which will be explored elsewhere.

\subsubsection{Inner Scale Lengths: $h_{R,i}$}
The absolute value of $h_{R,i}$ are plotted in the top row of Figure~\ref{h_brk} for the FUV, $V$, 3.6~$\mu$m, and H$\alpha$ data.  (Since the surface brightness for some of the FI and FI/II profiles {\it increases} with greater radius, $h_{R,i}$ is negative in these instances.  Consequently, $|h_{R,i}|$ is plotted to maintain the same axis ranges and allow the logarithmic plots.)  It is clear that more luminous dwarfs have larger inner disk scale lengths.  However, the inner scale lengths for the Type~III dwarfs are smaller than those of similarly luminous Type~I and II galaxies and generally bound between 0.1 and 3 kpc.  The $h_R$ values for the Type~I dwarfs are generally between those for Types~II and III.  The 1D K-S test probability that the dwarf Type~II and III $|h_{R,i}|$ values come from the same distribution is low: $1.8\times10^{-7}$.

As with Figure~\ref{Mu}, the small blue and red points in the $V$ panels of Figure~\ref{h_brk} represent 85 late-type spirals from PT06.  Since the scale lengths change very little between the $UBV$ bands, we have added to the $V$ panels the early-type $R$-band scale lengths from 66 barred spirals of EPB08 and 47 unbarred spirals of GEAB11.  The spiral inner scale lengths (late- through early-type) generally follow the same trends as the dwarfs with the $|h_{R,i}|$ values of Type~IIs being larger than those of IIIs and with the Type I values generally in the middle.  (See Table~\ref{log_trends} for the details of the fits.)  As with $\mu_{0,i}$, the PT06 spiral Types~II and III are still split in $h_{R,i}$ (with a K-S same distribution probability of $4.2\times10^{-7}$), and even more drastically when the early-type spirals are added, decreasing the K-S test probability to $1.5\times10^{-12}$.

Since any changes with wavelength are difficult to detect from Figure~\ref{h_brk}, the $|h_{R,i}|$, $h_{R,o}$, and $R_{br}$ values for four representative passbands (FUV, $U$, 3.6~$\mu$m, and H$\alpha$) are plotted as a function of the corresponding values in $V$ in Figure~\ref{Diff}.  Note that the samples change between passbands because not all galaxies were observed with all filters.  The upper row shows the results for $|h_{R,i}|$.  Interestingly, Type~IIs have shorter $|h_{R,i}|$ in the FUV than in $V$ and 3.6~$\mu$m whereas the reverse is true for Type~III profiles.  The FUV $y$-intercepts for the colored lines were determined by calculating the average of $\log_{10} |$FUV $h_{R,i}| - \log_{10} |V h_{R,i}|$.  A slope of unity was enforced because best fits initially allowing the slope to vary yielded slopes very close to unity for the larger samples, but the slope was unconstrained for the smaller samples.  Table~\ref{tabLogOffsets} lists the logarithmic offsets, the number of data points used for the fits, and the uncertainties in the mean for the passbands shown as well as for NUV, $B$, and $J$, all with respect to $V$.  Figure~\ref{Offsets} shows the logarithmic offsets for FUV, NUV, $UBJ$, and 3.6~$\mu$m.  The top panel shows the results for $|h_{R,i}|$.  Clearly the inner scale length {\it increases} with redder bands from the FUV to 3.6~$\mu$m for the Type~III profiles but {\it decreases} from the UV to the visible bands then is roughly constant in $J$ and 3.6~$\mu$m for the pure Type~II profiles.  Note that the Types FI and FI/II greatly increase the scatter in the {\it full} sample of Type~IIs.  The scatter in Figure~\ref{Diff} is also clearly larger for bands more separated in wavelength.

The scarcity of Type~I profiles in various bands precludes strong conclusions.  However, the top $U$ panel in Figure~\ref{Diff} shows that the Type~I $U$ scale lengths are similar but generally slightly smaller than those in $V$, more akin to the Type~III results than to the IIs.  The limited Type~I sample in Figure~\ref{Offsets} also shows a slight increase with wavelength.  Apparently the H$\alpha$ inner scale length (right most panel of Figure~\ref{Diff}) is generally larger than $|h_{R,i}|$ in $V$, with significant scatter.

\subsubsection{Outer Scale Lengths: $h_{R,o}$}
\begin{figure*}
\epsscale{0.98}
\plotone{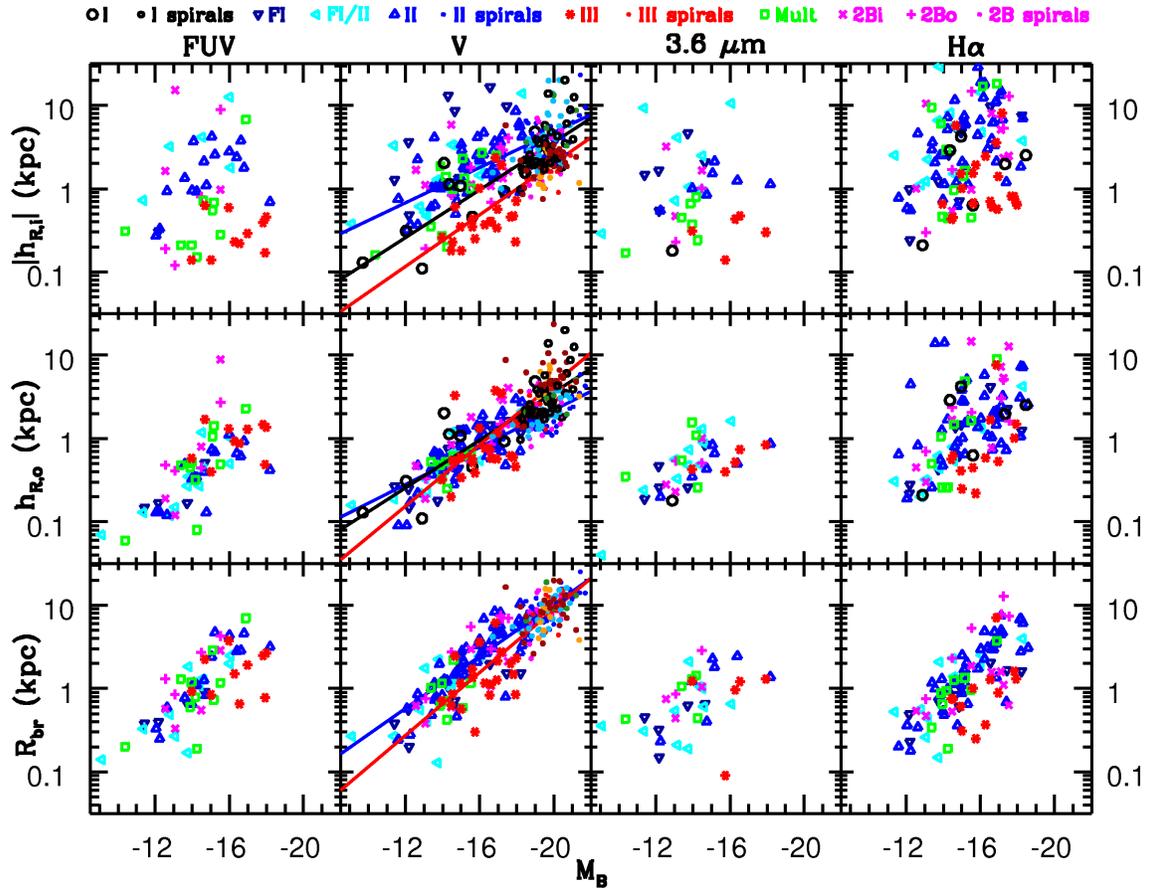}
\caption{Inner disk scale length, $|h_{R,i}|$ {\it (top)}, outer disk scale length, $h_{R,o}$ {\it (middle)}, and break location, $R_{br}$ {\it (bottom)} for four representative passbands revealing strong trends with $M_B$.  Note that the absolute value of $h_{R,i}$ is plotted because $h_{R,i}$ is negative for some Type~FI and FI/II profiles.  The single scale lengths for the Type~I profiles are plotted in the upper and middle rows for comparison.  The small blue (Type~II) and red (Type~III) spiral data plotted in the $V$ panels are actually $g'$ data from PT06.  Small sky blue (II) and orange (III) dots show $R$-band data for early-type barred spirals from EPB08 and likewise forest green (II) and maroon (III) for early-type unbarred spirals in $R$-band from GEAB11.  Note that Types~II and III are generally more separate in $|h_{R,i}|$ than in $h_{R,o}$ and the Type~II and III spirals are well mixed in $R_{br}$.  See Table~\ref{log_trends} for details about the least-square fits in the $V$ panels.  \label{h_brk} }
\end{figure*}

The middle row of Figure~\ref{h_brk} shows the outer scale lengths for the same four bands as a function of $M_B$.  There is a strong trend that more luminous galaxies have larger outer scale lengths.  Furthermore, all profile types have virtually collapsed on a single trend for both dwarfs and spirals, as seen in the middle $V$ panel, although the slope for Type~IIIs is still somewhat steeper.  (See Table~\ref{log_trends} for the details of the fits.)  The K-S test that $h_{R,o}$ for the different dwarf profile types follow the same distribution with $M_B$ give probabilities of: 41\% for Types~I and II, 26\% for Types~I and III, and 26\% for Types~II and III.  However, a close examination of the spiral data indicates that the Type~III spirals have larger outer scale lengths than the Type~II spirals such that the PT06 Types~II and III agree poorly (K-S same distribution probability of $1.9\times10^{-7}$) but more discrepancies arise when the early-type samples of EPB08 and GEAB11 are added (K-S same distribution probability of $2.5\times10^{-13}$).

\begin{deluxetable}{ccccc}
\tabletypesize{\scriptsize}
\tablecaption{Logarithmic Trend Fits of Scale Lengths and Break Location in $V$ from a combination of Dwarf and Spiral data \label{log_trends}}
\tablewidth{0pt}
\tablehead{ \colhead{Parameter} & \colhead{Type} & \colhead{Slope$^a$} & \colhead{y-int$^b$} & \colhead{Scatter$^c$} }
\startdata
$h_R$	    &	I    &	-0.143 &-2.30  &0.24 \\
$|h_{R,i}|$ &	II$^d$   &	-0.106$^e$ &-1.44  &0.24 \\
$|h_{R,i}|$ &	III  &	-0.155 &-2.79  &0.22 \\
$h_{R,o}$   &	II$^d$   &	-0.112$^e$ &-1.90  &0.17 \\
$h_{R,o}$   &	III  &	-0.184$^f$ &-3.02  &0.26 \\
$R_{br}$    &	II$^d$   &	-0.156 &-2.11  &0.17 \\
$R_{br}$    &	III  &	-0.188$^f$ &-2.82  &0.25
\enddata
\tablenotetext{$^a$}{The slope is $\log y$ (kpc) per $M_B$ magnitudes.}
\tablenotetext{$^b$}{The $y$-intercept is $\log y$ (kpc).}
\tablenotetext{$^c$}{The scatter is the standard deviation in the data when the trend is subtracted out.}
\tablenotetext{$^d$}{These statistics are for the sample of {\it pure} Type~IIs only.  The dwarf FI and FI/II (as well as two break and multiple profile) type samples have been ignored.}
\tablenotetext{$^e,^f$}{Note the similarities between these slopes, yielding constant ratios in Figure~\ref{Ratio}. }
\end{deluxetable}

\begin{figure*}
\plotone{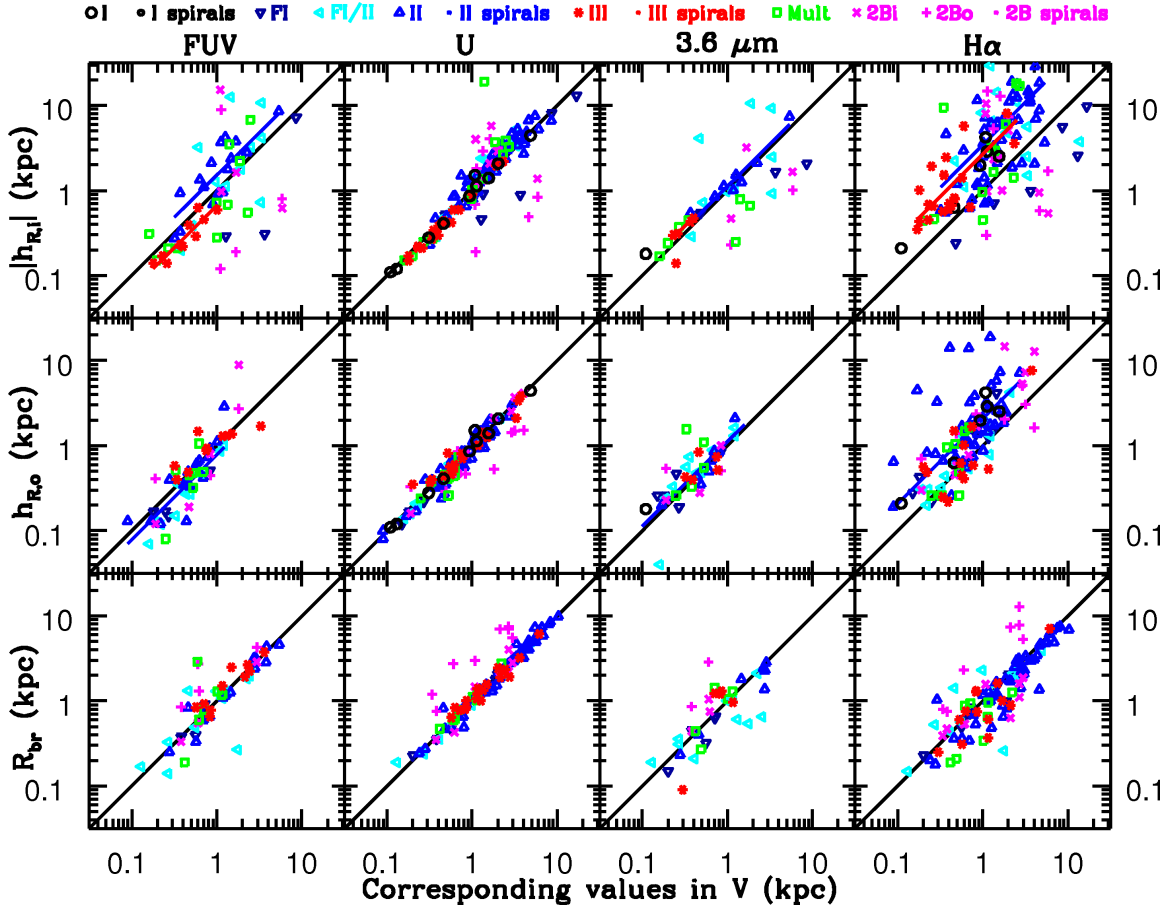}
\caption{Inner scale length, $|h_{R,i}|$ {\it (top)}, outer scale length, $h_{R,o}$ {\it (middle)}, and break location, $R_{br}$ {\it (bottom)} for four representative passbands as a function of the corresponding $V$ values.  Again, $h_R$ for the Type~I profiles are plotted in the upper and middle rows for comparison.  Each black line marks equal quantities.  The blue (Type~II) and red (Type~III) lines serve as guides for the color trends such that their offsets are the best determined fits under the condition of unity slope.  The Type~III and {\it pure} Type~II fits are shown for $|h_{R,i}|$ whereas fits for the {\it full} Type~II sample are shown for $h_{R,o}$.  Due to the crowded $U$ data around equality, no colored lines are shown in the $U$ panels.  Note that for Type~IIs, $|h_{R,i}|$ {\it decreases} for redder bands whereas the opposite is true for Type~IIIs.  However, for Type~IIs $h_{R,o}$ {\it increases} slightly for redder bands.  Also note that the break location is roughly independent of passband.  See Table~\ref{tabLogOffsets} for information about the $y$-intercepts for the best fit lines. \label{Diff} }
\end{figure*}

\begin{deluxetable*}{lccccccccccccccccccccc}
\tabletypesize{\scriptsize}
\tablecaption{Logarithmic Offsets for Scale Lengths and Breaks with respect to V \label{tabLogOffsets}}
\tablewidth{0pt}
\tablehead{
 & \multicolumn{3}{|c|}{FUV} & \multicolumn{3}{|c|}{NUV} & \multicolumn{3}{|c|}{U} & \multicolumn{3}{|c|}{B} & \multicolumn{3}{|c|}{J} & \multicolumn{3}{|c|}{3.6~$\mu$m} & \multicolumn{3}{|c|}{H$\alpha$} \\
\colhead{Profile} & \colhead{\#} & \colhead{Ave$^a$} & \colhead{$\sigma^b$} & \colhead{\#} & \colhead{Ave$^a$} & \colhead{$\sigma^b$} & \colhead{\#} & \colhead{Ave$^a$} & \colhead{$\sigma^b$} & \colhead{\#} & \colhead{Ave$^a$} & \colhead{$\sigma^b$} & \colhead{\#} & \colhead{Ave$^a$} & \colhead{$\sigma^b$} & \colhead{\#} & \colhead{Ave$^a$} & \colhead{$\sigma^b$} & \colhead{\#} & \colhead{Ave$^a$} & \colhead{$\sigma^b$}}
\startdata
I		&0  & \dots & \dots &  0 & \dots & \dots & 10 & -0.01 & 0.02 & 10 & -0.01  & 0.01  &  2 & 0.04  & 0.01 &  1 & 0.21  & \dots&  7 & 0.55  & 0.24 \\
II in$^c$	&33 & 0.23  &  0.09 & 35 & 0.31  & 0.08  & 80 & 0.33  & 0.06 & 81 & 0.32   & 0.05  & 20 & 0.28  & 0.14 & 23 & 0.30  & 0.13 & 80 & 0.72  & 0.07 \\
pII in$^d$&17 & 0.18  &  0.05 & 19 & 0.17  & 0.05  & 56 & 0.05  & 0.01 & 57 & 0.03   & 0.01  & 13 & -0.03 & 0.02 &  8 & 0.05  & 0.03 & 57 & 0.54  & 0.07 \\
III in	&10 & -0.18 &  0.03 & 10 & -0.12 & 0.02  & 19 & -0.05 & 0.01 & 21 & -0.03 & 0.01 & 12 & 0.04  & 0.02 &  4 & 0.08  & 0.01 & 21 & 0.43  & 0.06 \\
II out$^c$	&33 & -0.11 &  0.03 & 35 & -0.06 & 0.02  & 78 & -0.03 & 0.01 & 80 & 0.00   & 0.01  & 16 & -0.04 & 0.03 & 23 & 0.05  & 0.04 & 68 & 0.32  & 0.04 \\
pII out$^d$	&17 & -0.05 &  0.05 & 19 & -0.04 & 0.03  & 55 & -0.03 & 0.01 & 56 & 0.01   & 0.01  & 10 & -0.01 & 0.04 &  8 & 0.02  & 0.04 & 47 & 0.40  & 0.06 \\
III out		&10 & 0.06  &  0.06 & 10 & 0.02 & 0.06  & 19 & 0.00  & 0.02 & 21 & 0.00   & 0.02  &  7 & -0.03 & 0.03 &  5 & 0.04  & 0.07 & 13 & 0.09  & 0.08 \\
II br$^c$	&32 & -0.03 &  0.03 & 34 & -0.06 & 0.03  & 78 & -0.01 & 0.01 & 80 & -0.01 & 0.01 & 16 & -0.01 & 0.04 & 23 & -0.09 & 0.04 & 68 & -0.07 & 0.02 \\
pII br$^d$	&16 & -0.03 &  0.03 & 18 & -0.03 & 0.03  & 55 & -0.01 & 0.01 & 56 & -0.01 & 0.01 & 10 & -0.08 & 0.05 &  8 & -0.04 & 0.05 & 47 & -0.10 & 0.02 \\
III br		&10 & 0.05  &  0.03 & 10 & 0.05  & 0.03  & 19 & 0.00  & 0.02 & 21 & 0.00   & 0.01  &  7 & -0.02 & 0.04 &  5 & 0.00  & 0.14 & 13 & -0.09 & 0.06
\enddata
\tablenotetext{$^a$}{The logarithmic $y$-intercept with respect to $V$ of a line with unity slope determined from the best fit of shifting the unity slope line upwards or downwards.}
\tablenotetext{$^b$}{These are the uncertainties in the mean.}
\tablenotetext{$^c$}{These statistics are for the sample of {\it all} Type~IIs: pure IIs, FIs, and FI/IIs.}
\tablenotetext{$^d$}{Note that the pure~IIs (pII) agree much better with the full Type~II sample in $h_{R,o}$ and $R_{br}$ than in $|h_{R,i}|$.}
\end{deluxetable*}

\begin{figure}
\epsscale{1.0}
\plotone{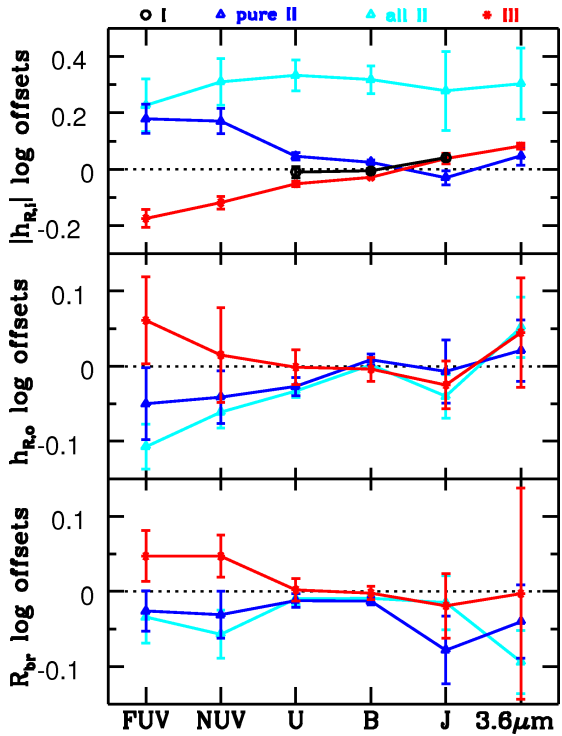} 
\caption{Wavelength trends in $|h_{R,i}|$ {\it (top)}, $h_{R,o}$ {\it (middle)}, and $R_{br}$ {\it (bottom)} for different profile types shown as logarithmic offsets with respect to $V$.   Figure~\ref{Diff} shows blue and red diagonal lines as fits to $|h_{R,i}|$, $h_{R,o}$, and $R_{br}$ in different bands as a function of the corresponding values in $V$.  Each colored line has a slope of unity but the offsets were determined for best fits.  These offsets are plotted here for the broad bands shown in Figure~\ref{Diff} as well as NUV, $B$, and $J$ and the numbers are listed in Table~\ref{tabLogOffsets}.  Note that $|h_{R,i}|$ {\it decreases} for redder wavelengths for pure Type~IIs but the reverse is true for Type~IIIs.  Also $h_{R,o}$ may increase for longer wavelength in Type~IIs whereas IIIs may have a slight opposing trend.  The break location is roughly invariant with wavelength for Types~II and III.  \label{Offsets} }
\end{figure}

Again, values of four passbands are plotted with respect to the $V$ values in Figure~\ref{Diff}, middle row.  Virtually no trends are detectable for the Type~III profiles and the data in the NUV and $BJ$ bands further support this consistency.  However, the outer scale lengths of Type~II profiles are slightly smaller in the FUV and slightly larger in the 3.6~$\mu$m data, indicating that the outer scale length may {\it increase} for redder bands, as detected by Zhang et~al.\,(2012) in a sample of 34 dwarfs heavily dominated by Type~IIs.  This is the opposite trend as the inner scale length of Type~II profiles and is also visible in the middle panel of Figure~\ref{Offsets}.  For $h_{R,o}$, this slight wavelength trend is seen in both the pure and full Type~II samples because the FIs and FI/IIs have $h_{R,o}$ values that generally overlap those of the pure~IIs.  However, this Type~II $h_{R,o}$ wavelength trend is weak; given the variations in the means, the data are marginally consistent with no wavelength trend at all.  Similarly from the middle panel of Figure~\ref{Offsets}, if there is a wavelength trend in the Type~III $h_{R,o}$ values, it opposes that of the Type~II $h_{R,o}$ values as well as the trend of the Type~III $h_{R,i}$ values.

A stronger trend appears in the middle H$\alpha$ panel of Figure~\ref{Diff}.  The Type~III dwarfs have $h_{R,o}$ values in H$\alpha$ that are similar to the $V$ values (99.9\% K-S chance of following the same distribution) whereas the Type~II H$\alpha$ $h_{R,o}$ values tend to be larger than the corresponding values in $V$, with only a $2.2\times10^{-4}$ K-S chance of following the same distribution.  Note, however, that the H$\alpha$ outer scale lengths are almost always less constrained than those in other bands because the H$\alpha$ data typically do not extend as far out radially as the other bands.  Lastly, the scale lengths of the outermost section of the dwarfs with two breaks (magenta cross-shaped points) appear to be smaller in $U$ and $B$ (not shown) than in $V$.

\subsubsection{Break Locations: $R_{br}$}
The break locations as a function of $M_B$ are plotted in the bottom row of Figure~\ref{h_brk}, again for FUV, $V$, 3.6~$\mu$m, and H$\alpha$.  Another clear trend exists such that more luminous galaxies have breaks farther out and the trends between Types~II and III are similar.  However, Type~III dwarf profiles break closer in (by roughly a factor of two) than Type~II dwarfs, again yielding a steeper slope with $M_B$ for IIIs and having a relatively low K-S probability (4.0\%) of coming from the same distribution.  The difference in the break location between Types~II and III disappears for spirals, with K-S probabilities of the same distribution of 92\% for the PT06 sample and 55\% for the full spiral sample (PT06, EPB08, GEAB11).

The bottom row of Figure~\ref{Diff} shows the break locations of the same four bands with respect to the $V$ values.  Again, the values for the second break appear slightly discrepant in $U$, $B$ (not shown), and H$\alpha$.  Otherwise, virtually no trends are detectable, so the break appears to occur at the same location regardless of passband.  The bottom panel of Figure~\ref{Offsets} further supports this conclusion since the logarithmic offsets with respect to $V$ are essentially zero for all six passbands.  (See Table~\ref{log_trends} for the numbers and uncertainties.)

\begin{deluxetable}{cccccc}
\tabletypesize{\scriptsize}
\tablecaption{Average Ratios between Scale Lengths and Break Location in $V$ \label{tabAveRatios}}
\tablewidth{0pt}
\tablehead{
 & & \multicolumn{2}{|c|}{Dwarfs} & \multicolumn{2}{|c|}{Spirals} \\
\colhead{Profile} & \colhead{Type} & \colhead{Ave} & \colhead{$\sigma$} & \colhead{Ave} & \colhead{$\sigma$}}
\startdata
$h_{R,o}/|h_{R,i}|$ &	II* &	0.47 &	0.02	 &	0.45 &	0.02     \\
$h_{R,o}/|h_{R,i}|$ &	III  &	1.85 &	0.20	 &	2.47 &	0.16     \\
$R_{br}/|h_{R,i}|$  &	II* &	1.48 &	0.11	 &	2.16 &	0.11     \\
$R_{br}/|h_{R,i}|$  &	III  &	3.21 &	0.19	 &	4.56 &	0.14     \\
$R_{br}/h_{R,o}$    &II* &	3.27 &	0.22	 &	4.77 &	0.17     \\
$R_{br}/h_{R,o}$    &III  &	1.93 &	0.16	 &	2.15 &	0.12
\enddata
\tablenotetext{*}{These statistics are for the sample of {\it pure} Type~IIs only.  The dwarf FI, FI/II, two break, and multiple profile type samples have been ignored.}
\end{deluxetable}

\begin{figure*}
\epsscale{0.98}
\plotone{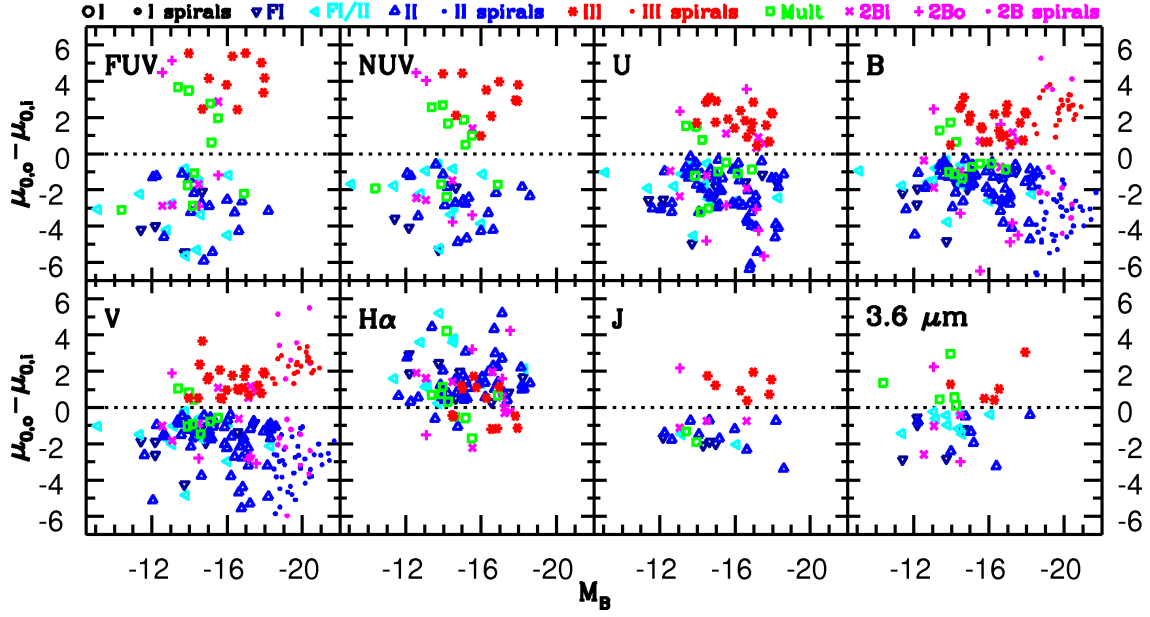}
\caption{Difference between outer and inner projected central surface brightness for the eight passbands with significant samples.  Again, the $V$-band data for the spirals were transformed from PT06 as indicated in the text (before Section 4.1.1) and here the $B$-band data for the spirals were transformed from PT06 using $B = g'+0.47(g'-r')+0.17$ (Smith et al.\,2002).  Also, the H$\alpha$ data are the logarithmic surface brightness instead of being in magnitudes and so are inverted with respect to the other panels.  The strength of the break clearly decreases for redder bands.  The spiral data suggest that breaks are stronger in brighter galaxies. \label{Out-In} }
\end{figure*}

\begin{figure*}
\epsscale{0.98}
\plotone{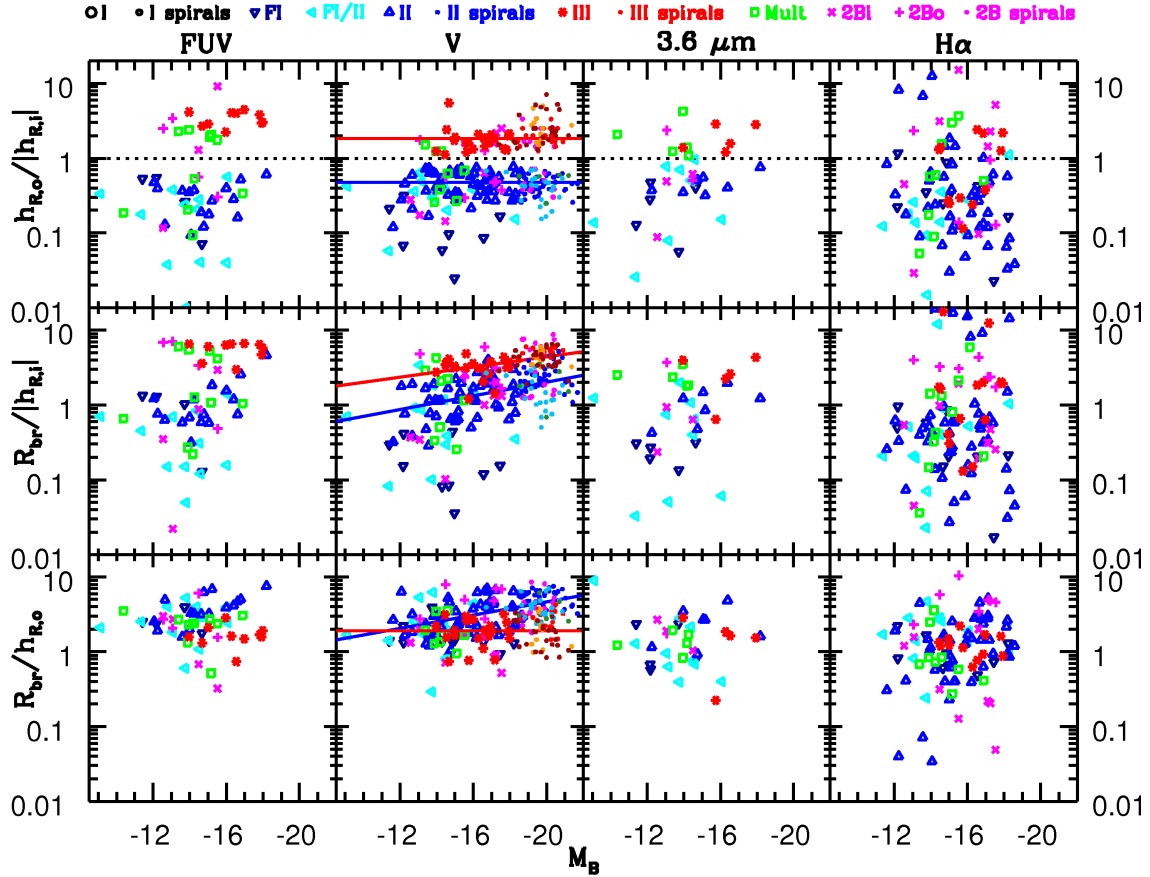}
\caption{Three ratios between the inner scale length, $|h_{R,i}|$, outer scale length, $h_{R,o}$, and break location, $R_{br}$, for four representative passbands.  See the caption of Figure~\ref{h_brk} for information about the small spiral dots.  Note that the ratios are relatively independent of $M_B$, and $V$-band average values for the dwarfs and spirals, separately, are given in Table~\ref{tabAveRatios}.  The one blue and two red horizontal lines in the $V$ top and bottom panels show the dwarf averages.  The two trends in the middle $V$ panels for pure Type~II and III profiles (dwarfs and spirals combined) are given by $\log (R_{br}/|h_{R,i}|) = -0.045 M_B - 0.60$ and $\log (R_{br}/|h_{R,i}|) = -0.034 M_B - 0.03$, respectively. The trend in the lower panel for pure IIs is given by $\log (R_{br}/h_{R,o}) = -0.044 M_B - 0.21$. \label{Ratio} }
\end{figure*}

\subsection{Inside versus Outside}
Several parameters can be examined to explore differences between the inner and outer regions of profiles with breaks.  Figure~\ref{Out-In} shows the difference between the outer and inner projected central surface brightness values in eight passbands.  Obviously, the Type~II and III profiles are separated due to the definitions of the broken profiles.  (The relationships are inverted for the H$\alpha$ panel again because the H$\alpha$ data are not in magnitudes but instead logarithmic surface brightness.  Additionally, the results are more mixed in H$\alpha$ than the other panels because the H$\alpha$ profiles did not always break the same way as the other bands.)  An interesting trend is that the break is clearly more apparent in bluer bands (i.e., the FUV) than redder bands.  While it is obvious that the Type~III (red points) are falling for redder passbands, the Type~II (blue points) are conversely rising for redder passbands.  Also, the small points in the $B$ and $V$ panels represent 85 late-type spirals from PT06.  Apparently the break is stronger in spirals than in dwarfs, at least in these two bands.

Figure~\ref{Ratio} explores several additional differences between the inner and outer regions and the parameters are quantized in Table~\ref{tabAveRatios}.  The top row of Figure~\ref{Ratio} shows results similar to those of Figure~\ref{Out-In}, except here the ratio between the outer and the inner scale length, the latter in absolute value, is being plotted.  Again, the Type~IIs and IIIs are generally separated with a greater distinction in the FUV than the other bands.  This is a result of the wavelength trends for $|h_{R,i}|$ and $h_{R,o}$.  Specifically, in the FUV, Type~II profiles have larger $|h_{R,i}|$ and smaller $h_{R,o}$ values with respect to $V$ so $h_{R,o}/|h_{R,i}|$ is especially low for FUV Type~IIs whereas Type~IIIs in the FUV have smaller $|h_{R,i}|$ and roughly equivalent $h_{R,o}$ values with respect to $V$ so $h_{R,o}/|h_{R,i}|$ is conversely elevated, explaining the larger gap in the FUV ratio.  Again, the H$\alpha$ results are somewhat mixed.

It is intriguing that the outer to inner scale length ratio varies with wavelength but is relatively constant with respect to $M_B$ for a given Type such that spirals have roughly the same $h_{R,o}/|h_{R,i}|$ values as dwarfs, at least in $V$, $g'$, and $R$ (see Table~\ref{tabAveRatios}).  The $h_{R,o}/|h_{R,i}|$ values for Type~III spirals are slightly higher than corresponding values for dwarfs, mainly from early-type spirals.

The middle and bottom rows of Figure~\ref{Ratio} show the ratios between the break radius and the inner and outer scale lengths, respectively.  Clearly the ratio of $R_{br}/|h_{R,i}|$ varies with wavelength and profile type because the inner scale length has clear trends with both variables whereas the break location is fairly invariant.  Additionally, $R_{br}/|h_{R,i}|$ increases slightly for more luminous galaxies for both Types~II and III.  The trends are especially apparent when both the dwarfs and spirals are considered.  It is not surprising that $R_{br}/|h_{R,i}|$ is generally larger for Type~IIIs than IIs since the inner scale lengths of Type~IIIs are generally smaller than those of IIs even though $R_{br}$ is also generally smaller for Type~IIIs than IIs or roughly equal for spirals.

As for $R_{br}/h_{R,o}$ (bottom row of Figure~\ref{Ratio}), since both the break radius and the outer scale length are fairly constant with wavelength (except for the slight trend in IIs), the ratio of $R_{br}/h_{R,o}$ is also relatively constant between passbands.  Furthermore, $R_{br}/h_{R,o}$ is relatively constant with $M_B$ for Type~IIIs due to the similar slopes in the $R_{br}$ and $h_{R,o}$ trends with respect to $M_B$ (see Table~\ref{tabAveRatios}).  However, $R_{br}/h_{R,o}$ is slightly higher for Type~II profiles than Type~III profiles in both dwarf and spiral galaxies and gradually increases from dwarfs to spirals.

\begin{deluxetable}{lcccc}
\tabletypesize{\scriptsize}
\tablecaption{Type Percentages of Four Studies\label{Tab4studies}}
\tablewidth{0pt}
\tablehead{           & \multicolumn{3}{|c|}{Spirals} & Dwarfs \\
\colhead{Profile} & \colhead{GEAB11} & \colhead{EPB08} & \colhead{PT06} & \colhead{Here\tablenotemark{$^a$}}\\
                               & \multicolumn{2}{c}{early-type} & late-type &  \\
                               & unbarred & barred & &}
\startdata
Type I					& 13 (28\%)	& 18 (27\%)	& 9 (11\%)	& 11 (8\%)	\\
Type II					&  5 (11\%)	& 28 (42\%)	& 45 (53\%)	& 86 (61\%)	\\
-- Pure II\tablenotemark{$^b$}	&  --			& --			& --			& 60 (70\%)	\\
-- Pure FI\tablenotemark{$^b$}	&  --			& --			& --			& 11 (13\%)	\\
-- FI/II\tablenotemark{$^b$}	&  --			& --			& --			& 15 (17\%)	\\
Type III					&  24 (51\%)	& 16 (24\%)	& 21 (25\%)	& 22 (16\%)	\\
Two Breaks				&  5 (11\%)	& 4 (6\%)		& 9 (11\%)	& 9 (6\%)		\\
-- III+II\tablenotemark{$^b$}	&  0 (0\%)		& 0 (0\%)		& 0 (0\%)		& 4 (44\%)	\\
-- II+III\tablenotemark{$^b$}	&  5 (100\%)	& 4 (100\%)	& 7 (77\%)	& 3 (33\%)	\\
-- II+II\tablenotemark{$^b$}	&  0 (0\%)		& 0 (0\%)		& 2 (22\%)	& 2 (22\%)	\\
Multiple\tablenotemark{$^c$}	& --			& --			& 1 (1\%)		& 13 (9\%)	\\
\tableline					
Total	&  47	& 66	& 85	& 141	\enddata
\tablenotetext{$^a$}{The profile type is for the galaxy as a whole over all the passbands, with the possible exception of H$\alpha$.}
\tablenotetext{$^b$}{Percentages are out of the total in the parent category.}
\tablenotetext{$^c$}{``Multiple'' indicates different profile types in different broadband passbands.}
\end{deluxetable}

\subsection{Type Percentages}
Large studies of spirals have found that most are better fit by broken, or double exponential, profiles than by a single exponential (PT06, EPB08, GEAB11).  We find this is also the case for our dwarf galaxies.  Out of 141 dwarfs, only 11 (8\%) have Type~I profiles in all the observed passbands, with the potential exception of H$\alpha$.  (Recall that the profile type of the H$\alpha$ data was frequently disregarded when determining the overall profile type of each galaxy.)  The majority, 86 (61\%), are some kind of Type~II followed by 22 (16\%) with Type~III profiles, again in all the observed broadband filters.  Of the remaining 22 profiles, 9 (6\% of the total) were best fit by two breaks and 13 (9\% of the total) have at least two profile types represented by the multiple passbands.  Note that these percentages are only approximate because some profiles were difficult to distinguish between Types~I, II, and III.  We consider galaxies with two breaks as a separate classification from galaxies with only one break.  In the literature, galaxies with two breaks are often included in the percentages of Type~II and III profiles with total percentages adding up to more than 100.

\begin{figure}
\plotone{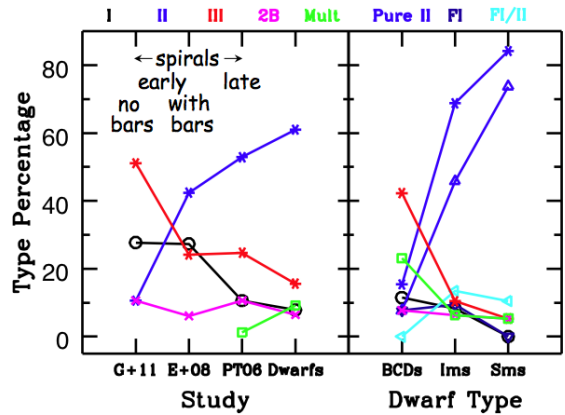}
\caption{Frequency of profile types in four studies.  In the right panel, the blue triangles show only the pure Type~II sample, whereas the blue asterisks include the pure FI and FI/II samples as well.  Note the trends with Hubble type such that later-type disks have more Type~II profiles and fewer Type~I and III profiles than early-type disks. \label{HubTrends}}
\end{figure}

Table~\ref{Tab4studies} lists the type statistics from this study as well as three others: PT06 focused on 85 late-type spirals in $g'$ and $r'$ whereas EPB08 and GEAB11 studied 66 and 47 early-type barred and unbarred spirals, respectively, in $R$.  As noted in GEAB11, some general trends exist between the frequency of profile types and galaxy Hubble type.  Specifically, later-type disks have more Type~II and fewer Type~I profiles.  Additionally, it appears that later-type disks have fewer Type~III profiles than early-type galaxies.  These trends are also shown in Figure~\ref{HubTrends}.

\subsubsection{Type II Subclassification}
In the above percentage break down, we have grouped several subclasses into the overall Type~II category.  Of the 86 Type~II profile galaxies, 11 (8\% total, 13\% II) are FI in all the considered passbands, 15 (11\% total, 17\% II) are a FI/II mixture, and 60 (43\% total, 70\% II) are purely Type~II.  Note that PT06 and EPB08 subclassify their Type~II profiles, but in a very different manner than what we have used here.  There are no analogs to FI profiles in spirals.

\subsubsection{Two Breaks}
Of the nine dwarf profiles that were best fit by two breaks in this study, four are III+II, three are some kind of II+III, and two are some kind of II+II where the inner and outer breaks are listed first then second, respectively.  (One of both the II+III and II+II categories has a compound FI/II as the inner break.)  As noted in Section 3.2.5 and broken down in Table~\ref{Tab4studies}, a few cases of III+II and II+II profiles have been identified in large samples of spiral galaxies (PT06, EPB08, GEAB11).  However, no III+II profiles have previously been presented and yet this mixed type is the most common in our sample of 141 dwarfs.  Also, the two early-type studies found no instances of II+II profiles whereas the two late-type studies found a total of four of these, indicating that Hubble type might be related to a double downbending profile.  Of course, all the two break galaxy considerations are severely limited by small number statistics.  The rarity of two break galaxies suggests that these mixed profiles perhaps only result from special circumstances.  The similar percentages from the early-type to late-type studies suggest no trend with Hubble type for two break galaxies, with the possible exception of double down-bending profiles.  Note that DDO~53 is the only galaxy in our sample with two breaks in the UV data but only one break in the other observed passbands.  However, the UV data of DDO~53 do probe significantly farther out than the other bands.

\subsubsection{Multiple Types for Different Passbands}
Since EPB08 and GEAB11 only explored $R$-band profiles, their studies do not contain information about breaks in different bands.  Using $g'$ and $r'$, PT06 presented one instance (IC~1067) with a Type~II $g'$ profile but a $r'$ profile barely consistent with being Type~I.  However, MacArthur et al.\,(2003) presented surface brightness profiles for 121 late-type spirals in the $BVRH$ bandpasses and used a ``transition'' class for galaxies whose profiles change from Type~II in the optical to Type~I in the infrared.  Consequently, a change of type with passband is not unprecedented.  Of the 13 dwarfs with multiple types for different passbands, 9 have $UBV$ profiles of the same type with a different profile type in the ultraviolet, near infrared, or mid infrared.  Two others have one profile type in $BV$ but another in $U$ and the ultraviolet.  Of the remaining two, UGC~199 has Type~II $UB$ but Type~I $V$ and NGC~1705 is lacking $UB$ data.  Apparently a change of type with passband generally requires a wider wavelength coverage than just $g'$ to $r'$.

\begin{deluxetable}{lcccc}[b]
\tabletypesize{\scriptsize}
\tablecaption{Type Percentages of Three Dwarf Categories and Galaxies with Eliminated Data\label{TabDwarfs}}
\tablewidth{0pt}
\tablehead{\colhead{Profile} & \colhead{dIms} & \colhead{BCDs} & \colhead{Sms} & \colhead{Eliminated}}
\startdata
Type I		& 8 (8\%)		& 3 (12\%)	& 0 (0\%)		& 5 (45\%)	\\
Type II		& 66 (69\%)	& 4 (15\%)	& 16 (84\%)	& 35 (41\%)	\\
-- Pure II		& 44 (67\%)	& 2 (50\%)	& 14 (88\%)	& 25 (42\%)	\\
-- Pure FI		& 9 (14\%)	& 2 (50\%)	& 0 (0\%)		& 2 (18\%)	\\
-- FI/II		& 13 (20\%)	& 0 (0\%) 		& 2 (12\%)	& 8 (53\%)	\\
Type III		& 10 (10\%)	& 11 (42\%)	& 1 (5\%)		& 5 (23\%)	\\
Two Breaks	& 6 (6\%)		& 2 (8\%)		& 1 (5\%)		& 3 (33\%)	\\
-- III+II		& 3 (50\%)	& 0 (0\%)		& 1 (100\%)	& 2 (50\%)	\\
-- II+III		& 1 (17\%)	& 2 (100\%)	& 0 (0\%)		& 0 (0\%)		\\
-- II+II		& 2 (33\%)	& 0 (0\%)		& 0 (0\%)		& 1 (50\%)	\\
Multiple		& 6 (6\%)		& 6 (23\%)	& 1 (5\%)		& 5 (38\%)	\\
\tableline
Total			& 96			& 26			& 19			& 53 (38\%)
\enddata
\tablenotetext{}{Note: Percentages of subcategories are out of the total in the parent category except for the Eliminated data column where the percentages are out of the total number of each Profile Type, i.e., column 5 of Table~\ref{Tab4studies}.}
\end{deluxetable}

\subsubsection{Dwarf Subclassifications: dIms, BCDs, \& Sms}
Of the 141 dwarfs, 96 are dIms, 26 are BCDs, and the remaining 19 are Sms.  Columns 2-4 of Table~\ref{TabDwarfs} show the percentages of the various profiles types for each of these three dwarf categories.  Since the dIms overwhelmingly dominate the full sample, it is not surprising that the dIm percentage breakdown of the various types closely follows those of the full sample.  Unfortunately, the relatively small sample sizes of BCDs and Sms preclude many conclusions.  However, almost half of the BCDs have Type~III profiles and 84\% of the Sms have Type~II profiles so at least these percentages are significantly different from the other two classes of dwarfs, respectively.

\subsubsection{Eliminated Data}
The last column of Table~\ref{TabDwarfs} lists the numbers (and percentages) of galaxies where at least one data point in at least one passband was eliminated as an outlier (see Section 3.2.3).  In general, the percentages by profile type are roughly equal to the total percentage (38\%).  The three possible exceptions are overabundances of outlying points in the Type~FI/II category (53\%) and underabundances in the pure FIs (18\%) and Type~IIIs (23\%). 

\begin{deluxetable}{lccc}
\tabletypesize{\scriptsize}
\tablecaption{Type Percentages of Dwarfs with \\Peculiar Morphology or Bars\label{TabPecBars}}
\tablewidth{0pt}
\tablehead{\colhead{Profile} & \colhead{Here\tablenotemark{*}} & \colhead{Pec Morph} & \colhead{Bars} }
\startdata
Type I		& 11 (8\%)	& 2 (18\%)	& 2 (18\%)	\\
Type II		& 86 (61\%)	& 14 (16\%)	& 23 (27\%)	\\
-- Pure II		& 60 (70\%)	& 12 (20\%)	& 18 (30\%)	\\
-- Pure FI		& 11 (13\%)	& 0 (0\%)		& 0 (0\%)		\\
-- FI/II		& 15 (17\%)	& 2 (13\%)	& 5 (33\%)	\\
Type III		& 22 (16\%)	& 3 (14\%)	& 6 (27\%)	\\
Two Breaks	& 9 (6\%)		& 2 (22\%)	& 2 (22\%)	\\
-- III+II		& 4 (44\%)	& 2 (50\%)	& 1 (25\%)	\\
-- II+III		& 3 (33\%)	& 0 (0\%)		& 0 (0\%)		\\
-- II+II		& 2 (22\%)	& 0 (0\%)		& 1 (50\%)	\\
Multiple		& 13 (9\%)	& 2 (15\%)	& 2 (15\%)	\\
\tableline
Total			& 141		& 23 (16\%)	& 35 (25\%)	
\enddata
\tablenotetext{*}{Same as column~5 of Table~\ref{Tab4studies} where percentages of subcategories are with respect to the total in the parent category.}
\tablenotetext{}{Note: The percentages in the third and fourth columns are with respect to the numbers in the second column.}
\end{deluxetable}

\subsection{Peculiar Morphology and Bars}
Of the 141 dwarf galaxies, 23 (16\%) were identified by \citet{he06} as having peculiar morphologies and 35 (25\%) were identified as being barred.  Table~\ref{TabPecBars} lists the number and percentages of each profile type (and subtype) with peculiar morphology (Column 3) and bars (Column 4).  Apparently there is no correlation between profile type and either peculiar morphology or the presence of a bar.

\begin{deluxetable*}{ccccc}
\tabletypesize{\scriptsize}
\tablecaption{Summary of $M_B$ Trends from Less to More Luminous Galaxies (i.e., $M_B << 0$)* \label{TabMB}}
\tablewidth{0pt}
\tablehead{\colhead{Parameter} & \colhead{Figure} & \colhead{II (blue $\triangle$)} & \colhead{III (red $\ast$)} & \colhead{Comparison between Types} }
\startdata
$\mu_{0,i}$	    &	8 (top)	   &	brighter   &	brighter  &	III brighter than II, Is in between \\
$\mu_{0,o}$	    &	8 (middle) &	brighter   &	constant  &	III generally dimmer than II, Is in between \\
$\mu_{br}$	    &	8 (bottom) &	$\sim$constant&	$\sim$constant &roughly 24 mag~arcsec$^{-2}$ \\
color$_i$	    &	9 (top)	   &	constant   &	constant  &	III bluer than II, Is closer to IIs \\
color$_o$	    &	9 (middle) &	constant   &	constant  &	III redder than II, Is in between \\
color$_{br}$	    &	9 (bottom) &	constant   &	constant  &	III redder than II \\
$|h_{R,i}|$	    &	10 (top)	   &	increases  &	increases &	III smaller than II, Is in between \\
$h_{R,o}$	    &	10 (middle) &	increases  &	increases &	Is, IIs, \& IIIs roughly overlap \\
$R_{br}$	    &	10 (bottom) &	increases  &	increases &	IIs \& IIIs roughly overlap \\
$h_{R,o}/|h_{R,i}|$ &	14 (top)   &	constant   &	slight increase&III higher than II by definition \\
$R_{br}/|h_{R,i}|$  &	14 (middle)&	increases  &	increases &	III higher than II \\
$R_{br}/h_{R,o}$    &	14 (bottom)&	increases  &	constant  &	III lower than II
\enddata
\tablenotetext{*}{Note that these trends are primarily determined from combining our dwarf $V$ results with spiral results from PT06 (in $g'$ and $r'$) and EPB08 and GEAB11 (both in $R$).}
\end{deluxetable*}

\begin{figure}
\epsscale{0.95}
\plotone{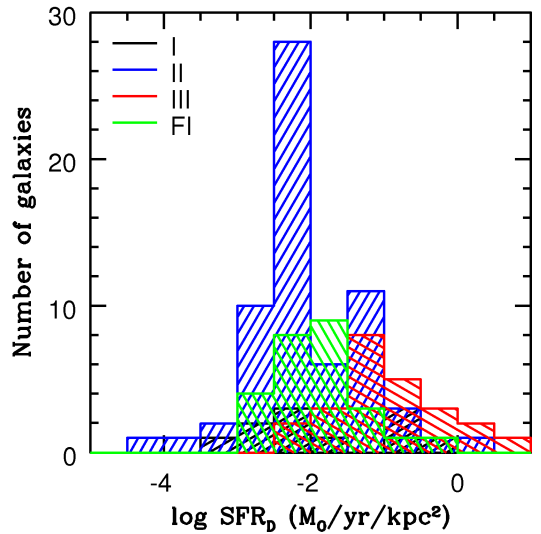}
\caption{Frequency of dwarf profile types as a function of H$\alpha$ SF rate (SFR) surface density.  Type~IIIs have systematically higher SFRs per unit area because they are predominantly BCDs, known to have high SFRs. \label{sfr_hist}}
\end{figure}

\subsection{Star Formation Rate}
Figure~\ref{sfr_hist} shows a histogram of the different profile types (in $V$) as a function of H$\alpha$ SFR surface density (normalized by the area within one disk scale length) from \citet{he04}.  Since the $V$-band profile type was used, the Type~FI/II samples and the multiple profile type samples are each represented in one of four main categories: I, II, III, or FI.  The small sample of nine two break galaxies were not plotted.  We see that Type~III galaxies systematically have higher SFR densities, but that is expected since Type~IIIs are primarily BCDs and BCD-like dIms and BCDs are known to have high SFRs (Hunter \& Elmegreen 2004 and references therein).  From a 1D two sample K-S test, the probabilities that the Type III samples follow the same distributions as the other profile types are 5.4\% (Is), $5.5\times10^{-7}$ (IIs), and $2.6\times10^{-5}$  (FIs).  However, there is a slight double peak for the Type~IIs such that the weaker peak at higher SFR corresponds to the peak of the Type~IIIs, with an 85\% probability that the Type IIIs and Type IIs with log~SFR$_D > -2$ follow the same distribution.  The peak of the FIs lies between the two Type~II peaks; the probability that log~SFR$_D$ of the FIs follows the same distribution as the IIs is small (1.4\%).

\subsection{Isolation}
Since the dwarf survey was initially chosen to be relatively isolated, most of the galaxies have tidal indices indicative of being relatively isolated, so nothing useful can be said about any possible relationship between current isolation and profile types.

\section{Summary and Discussion} 
Here we presented results from a human-assisted computerized method of fitting surface brightness profiles to 141 disk-like dwarf galaxies (96 dIms, 26 BCDs, and 19 Sms) in up to 11 passbands per galaxy.  Profiles that are well fit by a single line (or exponential) are Type~I whereas profiles requiring two lines (double exponential) with a steeper or shallower fall-off beyond the break are Type~II or III, respectively.  We also compare our dwarf results to three large studies of spirals: PT06, EPB08, and GEAB11.  We have found the following:

\begin{deluxetable}{cccc}[b]
\tabletypesize{\scriptsize}
\tablecaption{Summary of Wavelength Trends \\
from Bluer to Redder Wavelengths \label{TabLambda}}
\tablewidth{0pt}
\tablehead{\colhead{Parameter} & \colhead{Figure} & \colhead{II (blue $\triangle$)} & \colhead{III (red $\ast$)} }
\startdata
$\mu_{0,i}$	    &	8,9 (top)   &	brighter  &	brighter \\
$\mu_{0,o}$	    &	8,9 (middle)&	brighter  &	brighter \\	
$\mu_{br}$	    &	8,9 (bottom)&	brighter  &	brighter \\	
$|h_{R,i}|$	    &	11,12 (top)    &	$\downarrow$; steepens &	$\uparrow$; shallows \\		
$h_{R,o}$	    &	11,12 (middle)  &	$\sim\uparrow$; shallows &	$\sim$constant \\	
$R_{br}$	    &	11,12 (bottom)  &	constant  &	constant \\	
$h_{R,o}/|h_{R,i}|$ &	14 (top)    &	increases &	decreases \\	
$R_{br}/|h_{R,i}|$  &	14 (middle) &	increases &	decreases \\		
$R_{br}/h_{R,o}$    &	14 (bottom) &	decreases &	constant	
\enddata
\end{deluxetable}

\begin{figure*}
\epsscale{0.95}
\plotone{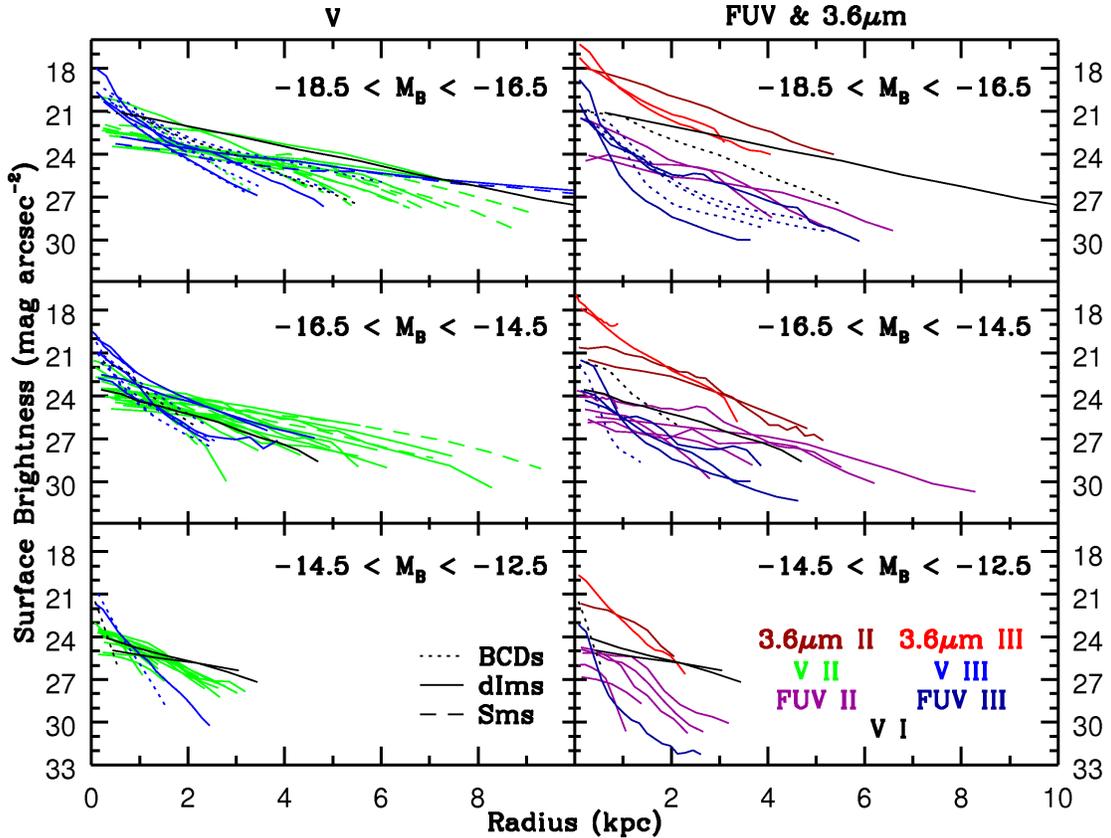}
\caption{Compilation of dwarf Type I, II, and III profiles for three different $M_B$ bins and in FUV, $V$, and 3.6~$\mu$m. Note that the axes are exactly the same for each panel and are in physical units: radius in kpc and the measured surface brightness values in mag~arcsec$^{-2}$.  The axes have not been scaled in any way to emphasize physical similarities and contrasts between the profiles.  The Type~I, II, and III profiles in $V$ are plotted on the left whereas Type II and III profiles in FUV and 3.6~$\mu$m are on the right, with the few Type~I $V$ profiles for comparison.  The inner and outer sections can be discriminated by eye via their change in slope.  Note (1) the differences between the Type II and III inner profiles, (2) the similarity between the Type II and III outer profiles, and (3) the difference in the break locations between IIs and IIIs.  These relationships are visible in all three bands, though the $V$ samples are significantly larger than the samples in the other two passbands.  Also note differences in the inner slope between passbands of Types III and II. \label{SumFig}}
\end{figure*}

\begin{packed_enum}
\item Clear trends exist between $M_B$ and fit parameters (see Table~\ref{TabMB}).  Brighter galaxies have brighter centers, larger inner and outer scale lengths, and break locations at larger radius.  All of these trends extend smoothly from dwarfs to spirals.
\item Several parameters are independent of $M_B$: (1) dwarfs have constant central colors; (2) $\mu_{br}$ in $V$ appears to be roughly constant at 24 mag~arcsec$^{-2}$ with a standard deviation of 1 mag~arcsec$^{-2}$ for dwarfs and spirals, regardless of profile type; (3) the ratios of $h_{R,o}/|h_{R,i}|$ and $R_{br}/h_{R,o}$ (the latter only for Type~IIIs) are essentially independent of $M_B$, though differ with passband and profile type.  (Again, see Table~\ref{TabMB}.) 
\item Dwarf Type~II and III profiles differ significantly with respect to their {\it inner} parameters ($\mu_{0,i}$ and $|h_{R,i}|$) but their outer ($\mu_{0,o}$ and $h_{R,o}$) parameters are similar, at least in the $V$-band.  By contrast, spiral Types~II and III are different in both inner and outer parameters.  Dwarf Type~IIs break roughly twice as far out in linear scale as dwarf IIIs but at roughly the same $\mu_{br}$ values whereas spiral Type~II and III break locations are very similar.
\item One difference in the outer regions of dwarf profiles is that $h_{R,o}$ values are equivalent in H$\alpha$ and $V$ in Type~IIIs whereas Type~II $h_{R,o}$ values are significantly larger in H$\alpha$ than in $V$ but slightly smaller in FUV than in $V$.  Is this due to a difference in (1) the structure of the disks or (2) the SF histories?  We will explore this further in a future paper.
\item For pure dwarf Type~IIs, the inner scale lengths {\it decrease} for redder bands (at least from FUV to $V$) whereas they {\it increase} for redder bands for dwarf~IIIs (see Table~\ref{TabLambda}).  The outer scale lengths for dwarf~IIIs are relatively constant (or barely decrease) from FUV to 3.6~$\mu$m and generally overlap the single scale length in Type~I profiles.  However, $h_{R,o}$ may increase for redder bands for all dwarf Type~II profiles, as seen by \citet{z+12}.  Even so, the break location is relatively constant with respect to wavelength for both Type~II and III dwarfs.
\item Breaks are stronger in the FUV than other bands for dwarfs and stronger in spirals than in dwarfs$-$ at least in $B$ and $V$, the only bands with possible spiral comparison.
\item Dwarf galaxies are consistent with late-type galaxies in trends between profile types and galaxy Hubble type such that later-type disk galaxies have more Type~IIs and fewer Type~Is and Type~IIIs than early-type disk galaxies.
\item BCDs and Sms are over-represented as Types~III and II, respectively, compared to dIms.
\item We present the first four reported galaxies with mixed surface brightness profiles of the Type~III+II (i.e., an up-bending break followed farther out by a down-bending break).  We also present 13 dwarfs with different profile types in different bands.
\item Bars and peculiar morphologies appear to have no influence on profile types.
\item Type~III dwarfs have higher SFR densities compared to other types because IIIs are dominated by BCDs which are known to have high SFRs.
\end{packed_enum}

Perhaps most remarkable is that a study of breaks in surface brightness profiles has any meaning at all in dwarf galaxies. These systems are generally irregular in appearance and SF is distributed stochastically in lumps that can be large relative to the size of the galaxy. Nevertheless, we find that these relatively isolated dIm, BCD, and Sm galaxies have azimuthally-averaged surface brightness profiles that are regular and exhibit an array of breaks similar to those seen in spirals. Furthermore, we have found regularity and trends in the break parameters, many of which continue into the regime of giant spiral galaxies. 

\begin{figure}
\plotone{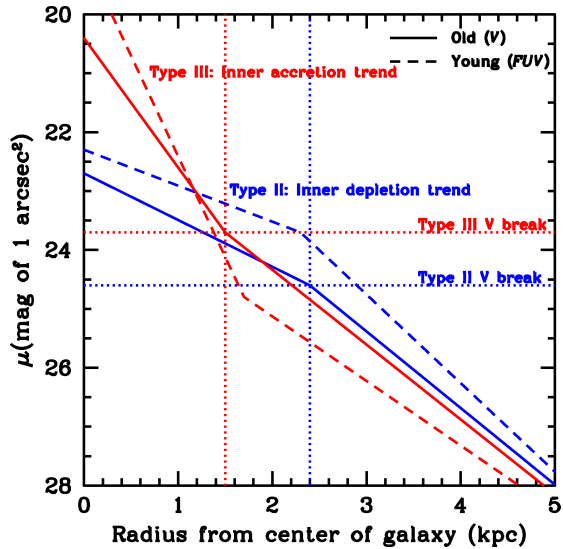} 
\caption{Representative $V$ and FUV (pure) Type~II and III profiles with parameters determined from averages or trend fits evaluated at $M_B=-16$.  Note (1) the different inner slopes between Types~II and III and between FUV and $V$ per profile type, (2) the similar outer slopes between Types~II and III, but slight shallowing trend from FUV to $V$ for IIs, (3) the similar break radii between FUV and $V$ but more distant break in Type IIs than IIIs, and (4) the similar break surface brightness values.  The steeper FUV slope of the Type~III centers implies an inner accretion trend whereas the shallower FUV slope of the Type~II centers corresponds to an inner depletion trend.  The steeper FUV slope in the Type~II outskirts is evidence of outside-in disk shrinking.  All magnitudes are on the Vega system. \label{Cartoon}}
\end{figure}

One key feature is that the surface brightness at the break radius in $V$ is generally the same over 10 mag in integrated galactic $M_B$. The standard deviation is roughly 1~mag, but the interesting point is that there is no trend with $M_B$ (see the bottom, second panel from the left, of Figure~8).  This may imply something fundamental about the relationship between the break radius and the stellar surface density, perhaps in turn implying a characteristic change in the SF process or efficiency at that radius. Furthermore, the strength of the break, as measured by the difference between the outer and inner projected central surface brightnesses, and the ratio of inner to outer disk scale lengths are also remarkably constant with $M_B$, again implying a regularity in the nature of the change at the break radius.  However, the surface brightness at the break varies from one band to another, so the constant $\mu_{br}$ in $V$ may result from competing influences, for example SF and mass redistribution.

Interestingly, these behaviors apply to both Type II and III breaks, even though we might expect different processes to be at play in producing the different profile types. In spite of these regularities, we do not see any relationship between break type and other galactic parameters, including the presence of bars or integrated SF rate except that BCDs, with their centrally concentrated SF activity, are more likely to be Type III than are dIm or Sm galaxies.

Another key feature is the comparison between Type II and III parameters for both dwarfs and spirals.  For dwarfs, the inner parameters are very different, the outer parameters are much more similar, and the Type II profiles break roughly twice as far out as the IIIs but at roughly the same surface brightness.  Figure~\ref{SumFig} summarizes these aspects for dwarf profiles by showing: (1) the Type~III inner profiles are much steeper than their Type~II counterparts, (2) the Type~II and III outer profiles are roughly parallel, and (3) the Type III profiles break closer in than the IIs.  Also, the inner Type III slope is steepest in the blue whereas the reverse is true for the inner Type II slopes, although it is not as pronounced.  Note that most BCDs have Type~III profiles but the Type III BCDs do not stand out specifically from the full Type III sample.  Consequently, it seems that all Type~III dwarfs are the same kind of beast where their inner steepening results from SF in their inner parts \citep{p+96, c+01, n+03}.  Perhaps BCDs are just the brightest or most concentrated population of Type III dwarfs.

Figure~\ref{Cartoon} illustrates these differences between Type~II and III dwarf profiles as well as their changes with passband. We have plotted representative $V$ and FUV profiles with parameters determined from averages or fits to trends from Tables~5-8 evaluated at $M_B=-16$.  The Type~III inner profile has a scale length that increases from the FUV to redder bands, so the slope of the profile is steeper in the FUV and the light falls off more quickly with radius than in $V$.  Conversely, for the Type~II inner profile, the slope is shallower in the FUV than in $V$.  The FUV is dominated by young stars formed over the past 200~Myrs, while $V$ is dominated by older stars formed over the past 1~Gyr for on-going SF.  Thus, these trends with wavelength suggest evolution over time, going from the solid lines to the dashed lines in the figure as a galaxy evolves.  The evolution of the profiles from $V$ to FUV for Type~II dwarfs suggests a trend of SF {\it depletion} in the inner disk, making the inner regions shallower.  For Type~III inner disks, the trend is just the opposite: increasing central SFR over time, possibly from {\it accretion}, which makes the inner regions steeper.  This interpretation for Type~IIIs is consistent with a history of interactions or minor mergers (e.g., Younger et~al.\,2007) that force a central concentration of gas and SF, particularly in BCDs which are usually Type~III.  Thus, the profile types and their trends with wavelength could be revealing differences in the evolutionary histories among dwarfs.

The outer regions do not change nearly as much with stellar population age or profile type as the inner regions.  Figure~\ref{Cartoon} hints that the outer slope is slightly steeper in the FUV than $V$ for the Type~IIs, but that the reverse is true for Type~IIIs.  These outer wavelength trends are based on data from Figure~\ref{Offsets} where the ``trends'' are very weak and the data are marginally consistent with no trend when the variations in the mean are taken into account.  Consequently, the outer trends in Figure~\ref{Cartoon} could be an artifact of the averaging process.  However, if the Type~II steeper outer FUV is real, it would be consistent with long term outside-in shrinking of the outer star-forming disk, as found by \citet{z+12}, and contrary to the inside-out paradigm for spirals.  If the Type~III shallower outer FUV is real, it would be inconsistent with the findings of \citet{z+12}; however, this is not surprising since their sample was dominated by Type~II dwarfs.

Between Type~II and III spirals, (1) the inner parameters are very different (like dwarfs), (2) the outer parameters differ (unlike dwarfs), but (3) the profiles break at roughly the same radius (also unlike dwarfs).  The difference between the dwarf and spiral outer profiles may be related to radial migration from interactions with spiral arms \citep{r+08a,r+08b,ms+09}, a mechanism that does not work in dwarfs.  Radial color trends should shed some light on this situation.

In subsequent papers, we will examine correlations between surface brightness breaks and color profiles, compare the breaks with gas surface density profiles, and look at 2D structure in the galaxies and its contribution to the breaks.  After examining these additional aspects, we will be in a better position to discuss what we have learned about the nature of surface brightness breaks and their causes in dwarf galaxies.

\acknowledgments
The authors would like to thank Elias Brinks, Hong-Xin Zhang, and Andreas Schruba for careful readings as well as the whole LITTLE THINGS team and the anonymous referee for helpful comments and suggestions.  This work is part of the LITTLE THINGS project, funded in part by the National Science Foundation through grants AST-0707563, AST-0707426, AST-0707468, and AST-0707835 to US-based LITTLE THINGS team members and with generous support from the National Radio Astronomy Observatory.  This research has made use of SAOImage DS9, developed by Smithsonian Astrophysical Observatory; the NASA/IPAC Extragalactic Database (NED) which is operated by the Jet Propulsion Laboratory, California Institute of Technology, under contract with the National Aeronautics and Space Administration; and NASA's Astrophysics Data System.

\end{document}